\documentclass[a4paper,11pt]{article}
\usepackage{amsmath,amsfonts,graphicx,yfonts,mathabx}
\usepackage[nosort]{cite}

\usepackage{xcolor}
\usepackage[colorlinks=true, linkcolor=black, urlcolor=blue, citecolor=blue]{hyperref}
\usepackage{amssymb}
\usepackage{epsfig, float}

\def\baselinestretch{1.1}
\newcommand{\be}{\begin{equation}}
\newcommand{\ee}{\end{equation}}
\newcommand{\bea}{\begin{eqnarray}}
\newcommand{\eea}{\end{eqnarray}}
\newcommand{\vev}[1]{{\left< {#1} \right>}}

\newcommand{\sac}{\, , \qquad}
\newcommand{\mt}[1]{\textrm{\tiny #1}}
\newcommand{\jem}{J^\mt{EM}}

\newcommand{\wn}{{\textswab{w}}}

\def\uh {u_\mt{H}}

\def\lgb {\lambda_\mt{GB}}

\renewcommand{\title}[1]{\vbox{\center\LARGE{#1}}\vspace{3mm}}
\renewcommand{\author}[1]{\vbox{\center#1}\vspace{3mm}}
\newcommand{\address}[1]{\vbox{\center\em#1}}
\newcommand{\email}[1]{\vbox{\center\tt#1}\vspace{3mm}}

\begin{document}

\begin{titlepage}
\begin{center}
\rightline{\tt}
\vskip 2.5cm
\title{Probing strongly coupled anisotropic plasmas from higher curvature gravity}
\vskip .6cm
\author{Viktor Jahnke, Anderson Seigo Misobuchi}
\vskip -.5cm 
\address{Instituto de F\'isica, Universidade de S\~ao Paulo\\ 05314-970 S\~ao Paulo, Brazil}
\vskip -.1cm 
\email{viktor.jahnke, anderson.misobuchi@usp.br}
\end{center}
\vskip 3cm

\abstract{
We consider five-dimensional AdS-axion-dilaton gravity with a Gauss-Bonnet term and use a black brane solution displaying spatial anisotropy as the gravity dual of a strongly coupled anisotropic plasma. We compute several observables relevant to the study of the plasma, namely, the drag force, the jet quenching parameter, the quarkonium potential, and the thermal photon production. The effects of higher derivative corrections and of the anisotropy are discussed and compared with previous results.}
\vfill
\end{titlepage}

\tableofcontents

\section{Introduction}

Results obtained at RHIC \cite{Adams:2005dq,Adcox:2004mh} and LHC \cite{lhc} suggest that  the quark-gluon plasma (QGP) produced in heavy ion collision experiments behaves as a strongly coupled system. The AdS/CFT correspondence \cite{duality1,duality2,duality3} has been a useful tool to understand such systems, since the duality maps a strongly coupled problem in the gauge theory into a relatively simpler problem in classical supergravity. There is an extensive literature regarding the applications of AdS/CFT to the physics of the QGP; see, e.g., \cite{CasalderreySolana:2011us} and references therein. However, most of these results rely on the strong assumption that $N_c$, the number of colors of the gauge theory, is infinite, as well as the 't Hooft coupling $\lambda=g_\mt{YM}^2N_c$. 
 
In order to describe more realistic gauge theories like QCD, it is crucial to investigate departures from the limit $N_c \to\infty$ and $\lambda \to\infty$. In the gravity dual side, this corresponds, respectively, to considering loop and higher curvature corrections to the supergravity action, which is a highly non-trivial task. In type IIB superstring, for instance, the leading order higher curvature correction arises as a term with schematic form $\alpha'^3R^4$ \cite{Gubser:1998nz}. Whatever the gravity dual of real world theories like QCD is, we will at some point have to deal with higher curvature corrections to go away from the infinite 't Hooft limit. Therefore, it is important to obtain as much information as regards the effect of higher curvature terms as possible. A simpler framework that allows us to study these higher curvature terms in a more tractable form is Lovelock gravity \cite{Lovelock:1971yv}. Lovelock theories can be thought of as a natural generalization of the Einstein-Hilbert action, constructed in such a way that the equations of motion are still second order in time derivatives so to avoid pathologies such as Boulware-Deser ghosts \cite{Boulware:1985wk}. Moreover, these theories admit a large class of asymptotically AdS black brane solutions which are particularly interesting in the context of the AdS/CFT correspondence.\footnote{Recent reviews about Lovelock gravity and its applications in AdS/CFT can be found in \cite{lovelock1,lovelock2,lovelock3}.} The simplest Lovelock theory includes a quadratic curvature term and is known as Gauss-Bonnet gravity. In view of the physics of the quark gluon plasma, a lot of attention was drawn to Gauss-Bonnet gravity since the discovery of a violation of the KSS bound \cite{Kovtun:2004de,Brigante:2007nu}. Some subsequent work regarding Gauss-Bonnet gravity in the study of the QGP includes \cite{Fadafan:2008gb, Fadafan1,Fadafan2,Fadafan3,Fadafan4,Fadafan5,Fadafan6,df15, chern}.

Another important aspect of the QGP is that at early stages after the collision it behaves as an anisotropic system due to the fact that the system expands mainly along the direction of the collision. A holographic model that incorporates the anisotropy was proposed by Mateos and Trancanelli in \cite{MT1,MT2} and subsequently many aspects related to it have been studied in detail. Some of the observables that have been studied include: the drag force \cite{Chernicoff:2012iq,Giataganas:2012zy}, the jet quenching parameter \cite{Giataganas:2012zy,Chernicoff:2012gu,Rebhan:2012bw}, the quarkonium potential and screening length \cite{Fadafan4,Giataganas:2012zy,Rebhan:2012bw,qqDiego}, the photon and dilepton production rate \cite{pp5,pp15}, the shear viscosity over entropy density \cite{shear1,shear2}, the Langevin diffusion coefficients \cite{langevin1,langevin2,langevin3}, and the addition of a chemical potential \cite{Cheng:2014qia,Cheng:2014sxa}; see \cite{Giataganas:2013lga} for a review of some of these computations.

In the present work, we explore the gravity solution obtained in \cite{Jahnke:2014vwa}  as the gravity dual of a strongly coupled anisotropic plasma. The anisotropy is incorporated by an axion field linear in one of the spatial directions, $\chi=az$, similarly to the model of Mateos and Trancanelli \cite{MT1,MT2}, and the higher curvature correction is given by the Gauss-Bonnet term
\begin{equation}
 \mathcal{L}_\mt{GB} = R^2-4 R_{mn}R^{mn} + R_{mnrs}R^{mnrs}.
\end{equation}
The action is given by
\begin{equation} \label{eq:action}
S=\frac{1}{16 \pi G}\int d^5x\, \sqrt{-g}\left[R+\frac{12}{\tilde{\ell}^2}-\frac{1}{2}(\partial \phi)^2-\frac{e^{2\phi}}{2}(\partial \chi)^2+\frac{\tilde{\ell}^2}{2}\lgb{\cal L}_\mt{GB}\right]+S_\mt{GH}\,,
\end{equation} 
where $\phi$ and $\chi$ are the dilaton and axion scalar fields, respectively, $S_\mt{GH}$ is a surface term necessary in order to have a well defined variational problem, $\tilde{\ell}$ is a parameter with dimension of length which we set to 1 in the following, and $\lgb$ is the (dimensionless) Gauss-Bonnet coupling. So we have two parameters in this model: the anisotropy parameter $a$ and the Gauss-Bonnet coupling $\lgb$. It is not obvious what is the combined effect of these two ingredients in the physical observables of the QGP, and our solution provides a simple framework where this effect can be investigated. Our philosophy is to consider simplified gravity models that, although not realistic, incorporate effects which are present in the quark gluon plasma. The idea is not to do precise phenomenological predictions but to understand qualitatively how different terms in the gravity theory affect the physical observables of the plasma.

One advantage of the above setup is that it allows us to obtain some results analytically, where the limits $a\to0$ and $\lgb\to0$ are under complete control. If $a \neq 0$ and $\lgb =0$, the dual field theory is known and corresponds to a deformation of the $\mathcal{N}=4$ SYM theory by a theta-term. On the other hand, the exact field theory dual to a Gauss-Bonnet gravity theory is not currently known, so our approach here is ``bottom-up'', where the parameter $\lgb$ plays the role of $\alpha'$, simulating $1/\sqrt{\lambda}$ corrections in the dual gauge theory. At least, we do know some aspects about the dual theory of Gauss-Bonnet gravity like, for example, that Gauss-Bonnet gravity is dual to a CFT with two different central charges \cite{aneqc,aneqc1,aneqc2}. It is also important to remember that, although in type IIB superstring the higher curvature correction is not of a Gauss-Bonnet type, the Gauss-Bonnet term may still appear as the $\alpha'$ correction of some other superstring theory as well. 

Our aim is to study qualitatively how several observables relevant to the study of the QGP are affected by the parameters $(a,\lgb)$. Anticipating the results of our analysis, we found that the effect of the Gauss-Bonnet term in the observables of the gauge theory is consistent with our physical intuition regarding the QGP as a fluid, interpreting the results in terms of the mean free path which is associated to the shear viscosity. Perhaps this consistency indicates that Gauss-Bonnet plays a role in the construction of a gravity theory dual to a realistic gauge theory like QCD.

This paper is organized as follows. In Section \ref{sec:sol} we review the gravity solution of \cite{Jahnke:2014vwa} and some thermodynamical properties such as the temperature and entropy. In Section \ref{sec:observables} we compute several observables relevant to study of the QGP, namely, the drag force experienced by a heavy quark moving through the plasma, the jet quenching parameter, the static potential between a quark-antiquark pair (quarkonium) and the photon production rate. We summarize and discuss our results in Section \ref{sec:discussion}. For completeness, the appendices contain the derivation of the main formulas used in this work.

\section{Gravity solution} \label{sec:sol}

\begin{figure}[H]
\begin{center}
\setlength{\unitlength}{1cm}
\includegraphics[width=0.97\linewidth]{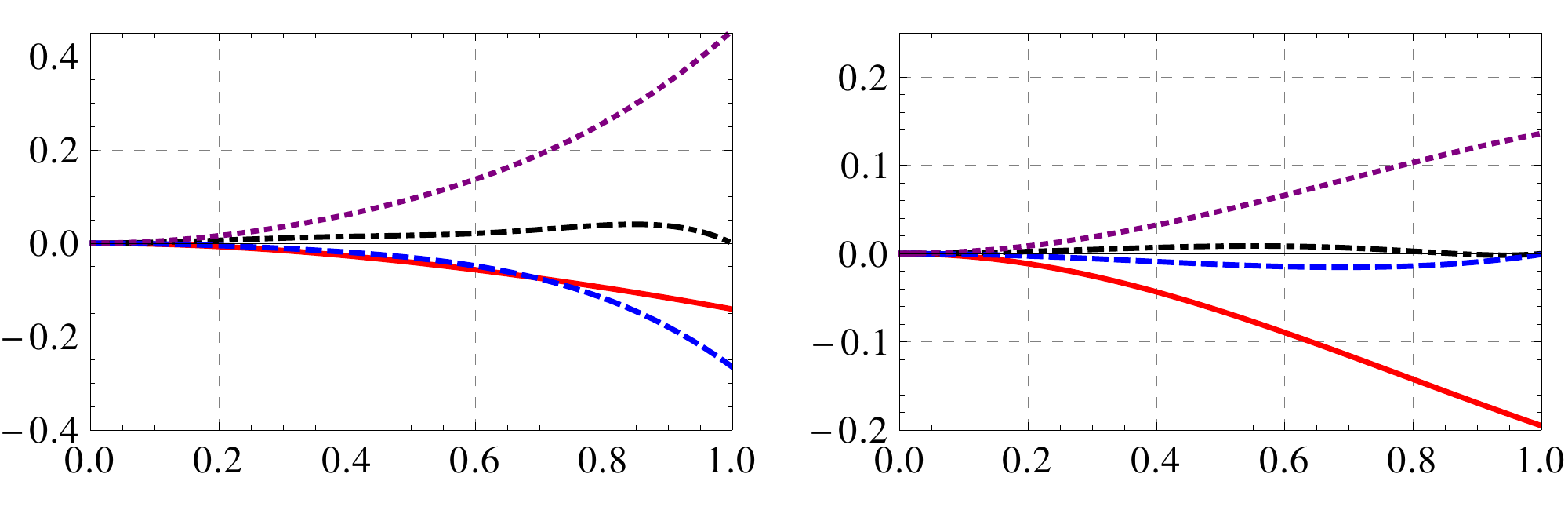} 
\put(-9.3,-.3){$u/\uh$}
\put(-9.3,4.2){$H_2$}
\put(-8.8,2.9){$F_2$}
\put(-8.7,1.2){$B_2$}
\put(-8.7,2.2){$\phi_2$}
\put(-1.4,-.3){$u/\uh$}
\put(-2.3,3.6){$H_2$}
\put(-2,2.7){$F_2$}
\put(-1.7,1.){$\phi_2$}
\put(-1.9,2.){$B_2$}
\end{center}
\caption{\small
The metric functions at order $O(a^2)$ for $\lgb=0.2$ (left) and $\lgb=-0.2$ (right).}
\label{fig-metric}
\end{figure}
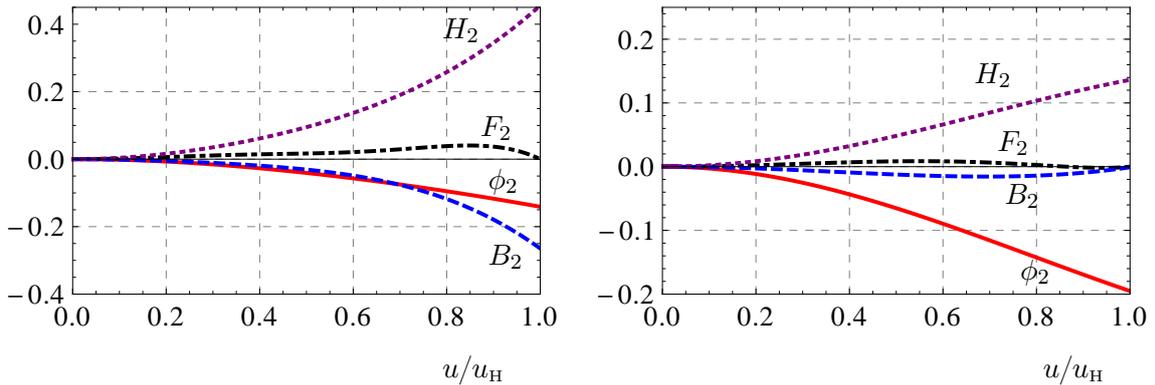

In this section we summarize the gravity solution obtained in \cite{Jahnke:2014vwa}. The Ansatz for the metric and scalar fields takes the form
\begin{gather}
ds^2=G_{mn}dx^mdx^n=\frac{1}{u^2}\left( -F(u) B(u)\, dt^2+dx^2+dy^2+H(u)\, dz^2 +\frac{du^2}{F(u)}\right),\nonumber \\
 \chi=az, \quad \phi=\phi(u).
 \label{eq:metric}
\end{gather}
The axion field introduces a spatial anisotropy in the $z$-direction controlled by the anisotropy parameter $a$. An analytical solution to the equations of motion derived from (\ref{eq:action}) can be obtained in the limit of small anisotropy. Their expressions take the form
\begin{align}
\phi(u)&=a^2 \phi_{2}(u)+O(a^4)\,,\cr
F(u)&=F_0(u)+a^2  F_{2}(u)+O(a^4)\,,\cr
B(u)&=B_0\left(1+a^2 B_{2}(u)+O(a^4)\right)\,,\cr
H(u)&=1+a^2  H_{2}(u)+O(a^4)\,.
\label{ansatz-sol}
\end{align}
The leading terms $F_0(u)$ and $B_0$ are known from pure Gauss-Bonnet gravity,
\begin{equation}
F_0(u)=\frac{1}{2\lgb}\left(1-\sqrt{1-4\lgb\left(1-\tfrac{u^4}{\uh^4}\right)}\right)\,,\quad B_0 =\frac{1+\sqrt{1-4\lgb}}{2}\,.
\end{equation}
The order $O(a^2)$ functions can be solved with the help of the auxiliary function
\begin{equation}
 U(u)\equiv\sqrt{1-4 \lgb  \left(1-\tfrac{u^4}{\uh^4}\right)},
\end{equation}
and the solution for them reads\footnote{The solution we present is written in a more compact form than in \cite{Jahnke:2014vwa}, but one can check that both forms are indeed equivalent.}
\begin{align}
 \phi_2(u)&=\frac{1}{8} \uh^2 \Big(\log \alpha(u)+U_0-U(u)\Big) \,,\cr
 F_2(u)&=\frac{u^4}{12 U_0^2 \uh^2 U(u)}\left(-\log \left(\tfrac{\alpha(u) }{\alpha(\uh)}\right)-\tfrac{6 u^4 \lgb}{\uh^4}+\tfrac{4 u^2\lgb}{\uh^2}+\tfrac{U_0^2 \uh^2}{u^2}+6 \lgb+U(u)-2\right)\,,\cr
  B_2(u)&=\frac{u^2 \uh^2}{24 U_0^2 \left(\uh^2+u^2\right)}\left(-\tfrac{6 u^4 \lgb}{\uh^4}-\tfrac{2 u^2 \lgb}{\uh^2}+4 \lgb-\tfrac{8 \phi_2(u)}{\uh^2}-\tfrac{8 \phi_2(u)}{u^2}-2 U(u)-2\right),\cr
   H_2(u)&=\frac{\uh^2}{8 U_0^2} \left(-\log \alpha(u) +2\lgb\tfrac{u^2}{\uh^2}\left(\tfrac{u^2}{\uh^2}-2\right)+U(u)-U_0\right) \,,
\end{align}
where $U_0\equiv U(0)$ and
\begin{equation}
 \alpha(u)=\left(\frac{2 u^2 \sqrt{\lgb}+\uh^2 U(u)}{U_0 \uh^2}\right)^{2 \sqrt{\lgb}}\frac{ \left(-4 \lgb \uh^2 \left(\uh^2+u^2\right)+\uh^4 U(u)+\uh^4\right)}{(2 B_0-4 \lgb) \left(\uh^2+u^2\right)^2} \,.
\end{equation}
The boundary conditions were fixed such that $F$ vanishes at the horizon $u=\uh$. At the boundary $u=0$ we have
\begin{equation}
 \phi_{2,\mt{bdry}}=0, \ \  F_{2,\mt{bdry}}=0, \ \  B_{2,\mt{bdry}}=0, \ \  H_{2,\mt{bdry}}=0.
\end{equation}
The solution is regular everywhere and asymptotically approaches $AdS_5$. A plot of the metric functions is shown in Fig. \ref{fig-metric}.
Thermodynamical quantities such as temperature and entropy can be computed via standard formulas, by requiring the regularity of the (Euclideanized) metric at the horizon and using the area law, respectively. One finds
\begin{align}
T&=\sqrt{B_0}\left(\frac{1}{\pi\uh}-\frac{2B_0-6\lgb+\sqrt{\lgb}\log\left(\frac{1+2\sqrt{\lgb}}{1-2\sqrt{\lgb}}\right)
-\log\left(\frac{4B_0}{\sqrt{1-4\lgb}}\right)}{48\pi(1-4\lgb)}\uh a^2+O(a^4)\right) \nonumber\,, \\
s& =\frac{\pi}{4GB_0^{3/2}}\left(\pi^2 T^3+\frac{1}{8}T B_0 a^2+O(a^4)\right)\,.
 \label{eq:temp-s}
\end{align}

\section{Observables in a strongly coupled anisotropic plasma} \label{sec:observables}

In this section we compute several observables relevant to the study of the QGP. Most part of the analysis of our results involves a comparison with the isotropic $\mathcal{N}=4$ SYM result, obtained by taking $a\to0$ and $\lgb\to0$. We limit ourselves to the comparison at the same temperature, for simplicity, but a comparison at the same entropy density is still possible, and it was done for the observables computed in the model of \cite{MT1,MT2}. Some quantities were already computed in \cite{Jahnke:2014vwa}, such as the shear viscosity over entropy density $\eta/s$ and the electric conductivities.

\subsection{Drag force} \label{subsec:drag}

When a heavy quark propagates through a strongly coupled plasma it loses energy due to the interaction with the medium. One quantity related to the dissipation of energy of the quark is the drag force. The study of drag force in a strongly coupled plasma was initiated in \cite{Gubser:2006bz,Herzog:2006gh} for the case of (isotropic) $\mathcal{N}=4$ SYM and subsequently it was extended in several ways. See, for instance \cite{df15,Giataganas:2012zy,df1,df2,df3,df4,df5,df6,df7,df8,df9,df10,df11,df12,df14,df16,df17,df18,df19,df20}. The two computations of the drag force closely related to the present work were done in \cite{Fadafan:2008gb,Chernicoff:2012iq}, corresponding to the limits $\lgb=0$ and $a=0$, respectively.

Following the standard prescription of the computation of the drag force, we consider an external heavy quark moving through the strongly coupled plasma with constant velocity $v$. Since the heavy quark loses energy due to the interaction with the plasma, an external force is necessary to maintain the motion stationary. In the dual picture, we have a classical string with an endpoint in the quark (at the boundary) and the other endpoint in the bulk, in a picture usually referred to as ``trailing string'' \cite{Gubser:2006bz,Herzog:2006gh}. The derivation of the general formula is presented in Appendix \ref{app:drag}. As a result, we first need to determine a critical point $u_c$ by solving the equation 
\begin{equation}
 \left[\frac{2G_{tt}}{v^2}+G_{xx}+G_{zz}+(G_{zz}-G_{xx})\cos(2\varphi)\right]_{u=u_c}=0,
  \label{eq:ucritical}
\end{equation}
where $\varphi$ is the angle between the direction of motion of the quark and the $z$-direction. In the following, we will be interested in the cases where the motion of the quark is parallel ($||$) and transversal ($\perp$) to the direction of anisotropy, corresponding to $\varphi=0$ and $\varphi=\pi/2$, respectively.  Once the critical point is determined, it is straightforward to compute the drag force using
\begin{equation}
 F_\mt{drag}^{\,||}= e^{\phi/2} G_{zz}v\Big|_{u=u_c} ,\qquad  
 F_\mt{drag}^{\perp}= e^{\phi/2} G_{xx}v\Big|_{u=u_c}.
  \label{eq:dragformula}
\end{equation}
Since we are working in the small anisotropy regime, the critical point can be written as
\begin{equation}
  u_c={{u_ 0}_c}+a^2{{u_ 2}_c}+O(a^4).
\end{equation}
For our particular background (\ref{eq:metric}), the equation for the critical point (\ref{eq:ucritical}) expanded to second order in $a$ becomes
\begin{equation}
B_0 F_0 -v^2 + a^2 \left(B_0 B_2 F_0 + B_0 F_2 -v^2 \cos^2\varphi H_2 + B_0 {u_2}_c F_0'\right) \Big|_{{u_0}_c}=0,
\end{equation}
Solving the equation order by order gives
\begin{align} \label{eq:ucsol}
{{u_ 0}_c} & =\uh\left(\frac{B_0^2-v^2 B_0+v^4 \lgb}{B_0^2}\right)^\frac{1}{4}, \nonumber \\
{{u_ 2}_c} & =-\frac{B_0 B_2({{u_0}_c}) F_0({{u_0}_c})+B_0 F_2({{u_0}_c})-v^2 H_2({{u_0}_c}) \cos ^2\varphi }{B_0 F_0'({{u_0}_c})}.
\end{align}
Plugging the solution for the critical point (\ref{eq:ucsol}) into the formulas of the drag force (\ref{eq:dragformula}), we obtain
\begin{align}
F_\mt{drag}^{\,||} & =\frac{v }{{{u_0}_c}^2}+\frac{a^2 v}{2 {{u_0}_c}^2} \left(\phi_2({{u_0}_c}) -4 \frac{{{u_2}_c}}{{u_0}_c}+2 H_2({{u_0}_c}) \right) + O(a^4),\nonumber\\
F_\mt{drag}^{\perp} & =\frac{v}{{{u_0}_c}^2} + \frac{a^2 v }{2 {{u_0}_c}^2}\left(\phi_2({{u_0}_c})-4 \frac{{{u_2}_c}}{{u_0}_c}\right) + O(a^4).
\end{align}
We do not report the full explicit expressions for the drag force here since they are too long and not very illuminating. Inverting the first relation of (\ref{eq:temp-s}), the drag force can be expressed as a function of the temperature.\footnote{The easiest way to write the drag force in terms of the temperature is by noting that the critical point scales as $u_c=\uh\gamma_0+a^2\uh^3\gamma_2+O(a^4)$, where $\gamma_0$ and $\gamma_2$ are quantities that do not depend on $\uh$.} We can then check that in the limit $\lgb\to0$ we recover the result of \cite{Chernicoff:2012iq},
\begin{align}
 F_\mt{drag}^{||\,\mt{MT}} & =\frac{\pi ^2 T^2 v}{\sqrt{1-v^2}} +\frac{a^2 v \left(-v^2+\sqrt{1-v^2}+\left(v^2+1\right) \log \left(\sqrt{1-v^2}+1\right)+1\right)}{24 \left(1-v^2\right)^{3/2}}, \nonumber \\
 F_\mt{drag}^{\perp\,\mt{MT}} & =\frac{\pi ^2 T^2 v}{\sqrt{1-v^2}}+\frac{a^2 v \left(-v^2+\sqrt{1-v^2}+\left(4 v^2-5\right) \log \left(\sqrt{1-v^2}+1\right)+1\right)}{24  \left(1-v^2\right)^{3/2}},
\end{align}
and in the limit $a\to0$ we recover the result of \cite{Fadafan:2008gb}
\begin{equation}
F_\mt{drag}^{\,\mt{GB}}= \frac{\sqrt{2} \pi ^2 T^2 v}{\sqrt{\left(v^2-1\right) \left(2 \left(v^2+1\right) \lgb-\sqrt{1-4\lgb}-1\right)}}.
\end{equation}
Of course, in the limit where both $a$ and $\lgb$ go to zero we recover the isotropic $\mathcal{N}=4$ SYM result \cite{Gubser:2006bz,Herzog:2006gh}
\begin{equation} \label{eq:dragiso}
 F_\mt{drag}^{\,\mt{iso}} =\frac{\pi ^2 T^2 v}{\sqrt{1-v^2}}.
\end{equation}

\begin{figure}[H]
\begin{center}
\setlength{\unitlength}{1cm}
\includegraphics[width=0.97\linewidth]{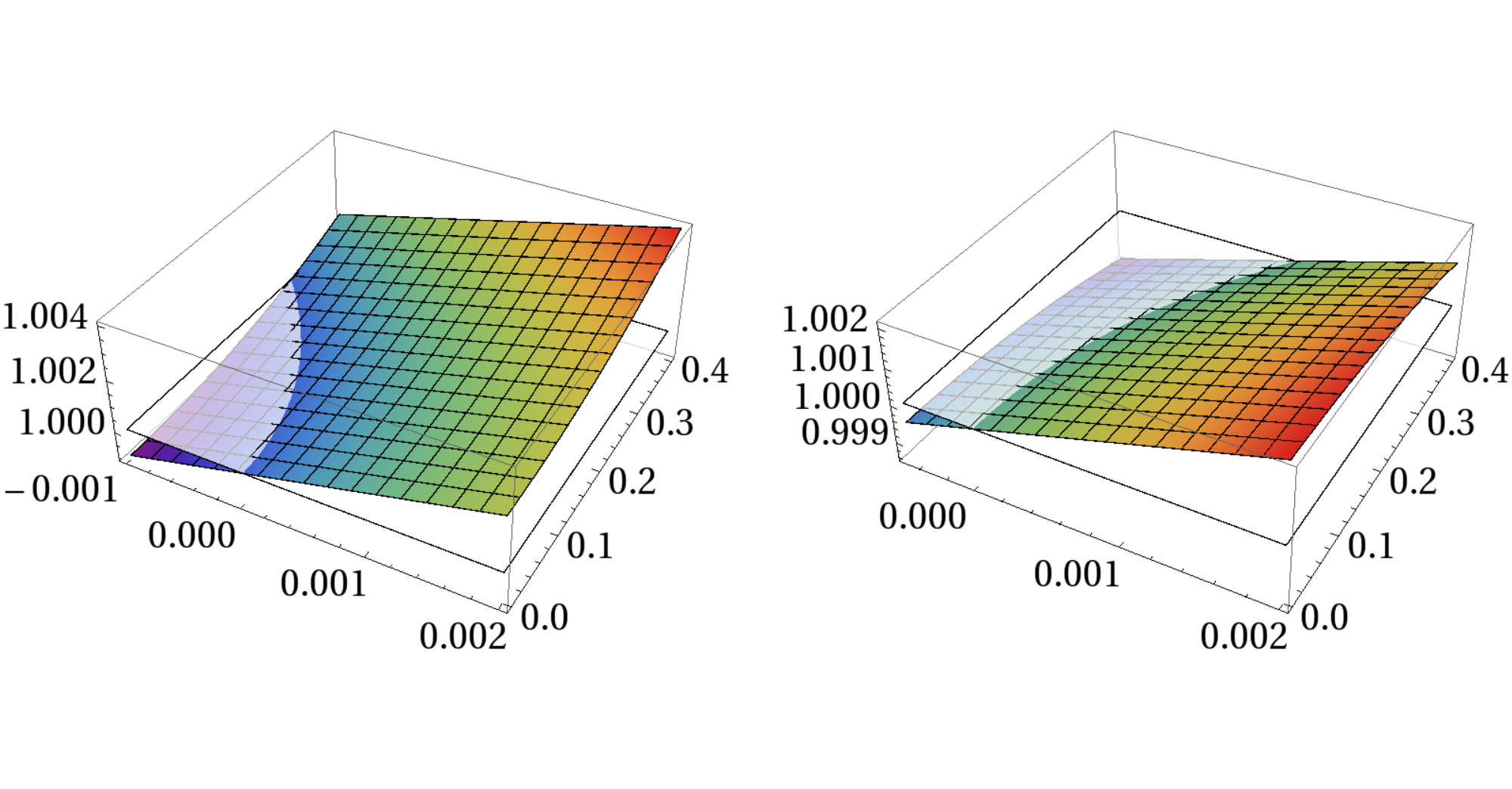} 
  \put(-12.0,6.8){$F_\mt{drag}^{\,||}/F_\mt{drag}^{\,\mt{iso}} $}
  \put(-4.2,6.8){$F_\mt{drag}^{\perp}/F_\mt{drag}^{\,\mt{iso}} $}
  \put(-12.7,1.5){$\lgb$}
  \put(-4.7,1.5){$\lgb$}
   \put(-8.7,2.5){$a/T$}
  \put(-0.7,2.5){$a/T$}
\end{center}
\vspace{-30pt}
\caption{Drag force normalized by the isotropic result as a function of $(\lgb,\frac{a}{T})$. Here we have fixed $v=0.3$. Left: Motion along the anisotropic direction. Right: Motion along the direction transversal to the anisotropy.}
\label{fig:drag1}
\end{figure}

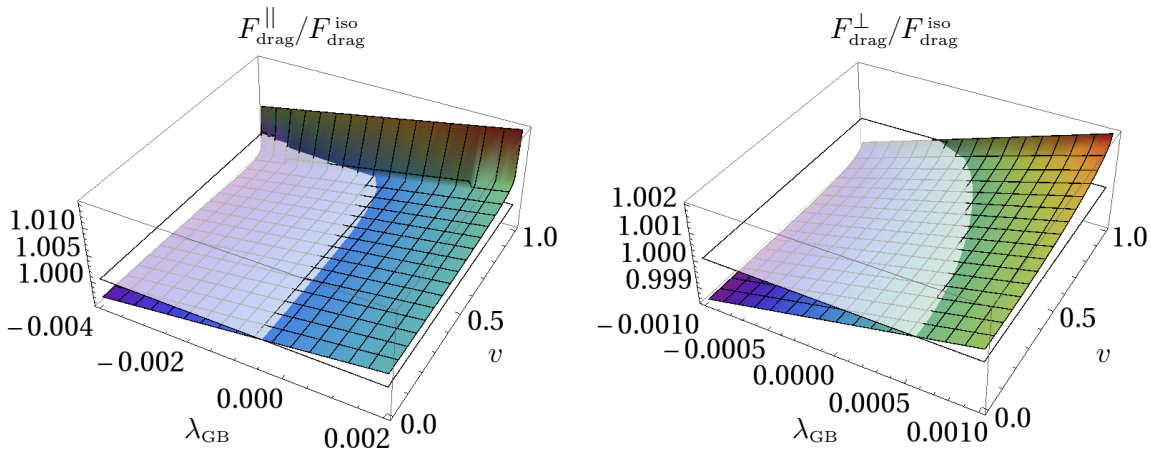
\begin{figure}[H]
\begin{center}
\setlength{\unitlength}{1cm}
\includegraphics[width=0.97\linewidth]{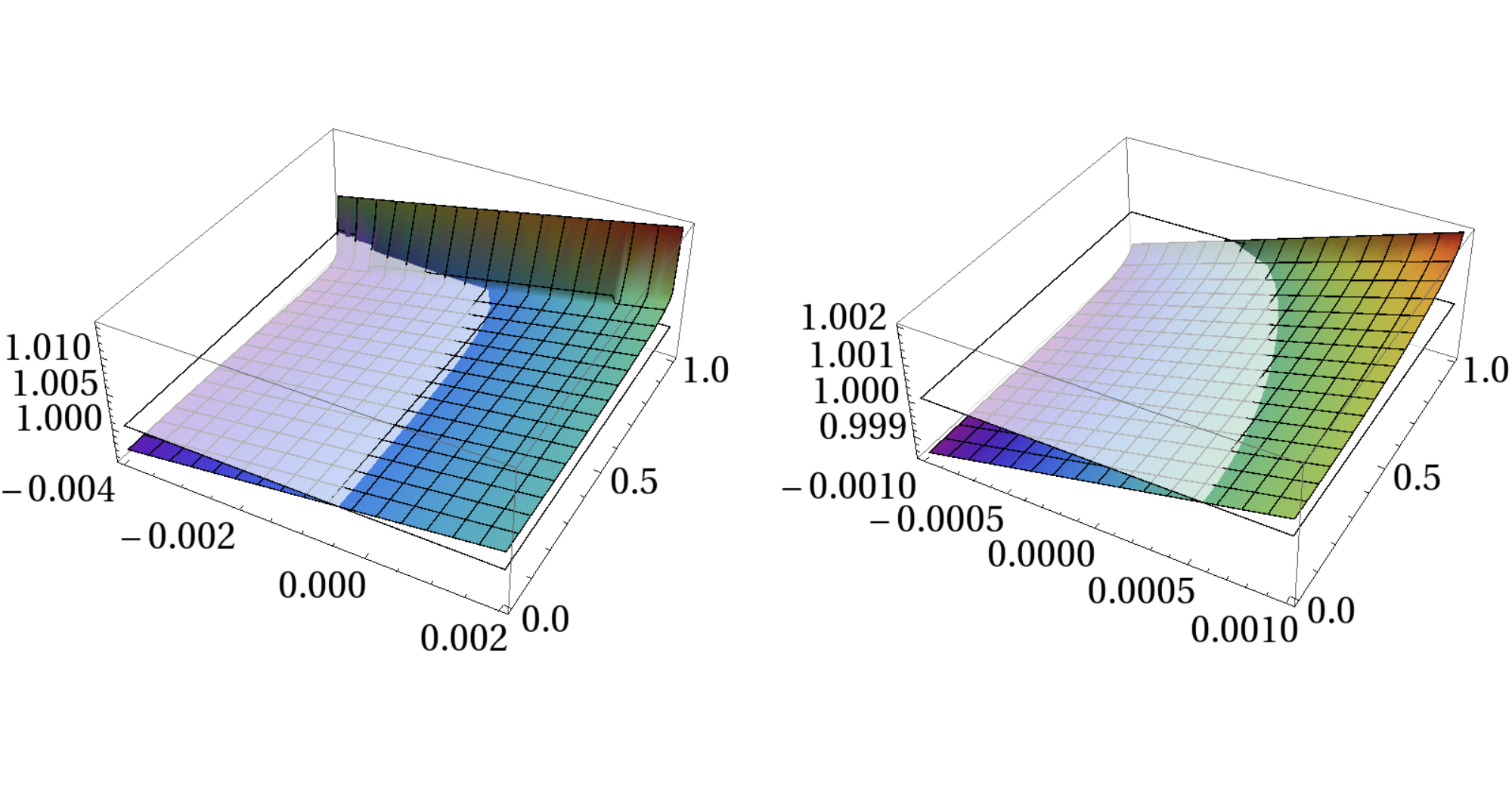} 
  \put(-12.0,6.8){$F_\mt{drag}^{\,||}/F_\mt{drag}^{\,\mt{iso}} $}
  \put(-4.2,6.8){$F_\mt{drag}^{\perp}/F_\mt{drag}^{\,\mt{iso}} $}
  \put(-12.7,1.5){$\lgb$}
  \put(-4.7,1.5){$\lgb$}
   \put(-8.7,2.5){$v$}
  \put(-0.7,2.5){$v$}
\end{center}
\vspace{-30pt}
\caption{Drag force normalized by the isotropic result as a function of $(\lgb,v)$. Here we have fixed $\frac{a}{T}=0.2$. Left: Motion along the anisotropic direction. Right: Motion along the direction transversal to the anisotropy. For other values of $\frac{a}{T}$ the results were qualitatively the same.}
\label{fig:drag2}
\end{figure}

In the analysis of our results, it is useful to normalize the drag force by the isotropic result (\ref{eq:dragiso}). The normalized drag force here depends on $v, \lgb$ and $a/T$. The main result is shown in Fig. \ref{fig:drag1}. Our results are, as expected, a combined effect of their limiting cases \cite{Chernicoff:2012iq,Fadafan:2008gb}. The effect of the Gauss-Bonnet coupling is, in general, to enhance the drag force for $\lgb>0$ and to decrease it for $\lgb<0$, for both longitudinal and transversal motion. This is the same effect observed in the case of pure Gauss-Bonnet gravity \cite{Fadafan:2008gb}, but it is different from what happens with corrections of type ${\alpha'}^3R^4$, where the drag force is always enhanced \cite{df19}. The effect of the anisotropy is qualitatively the same found in  \cite{Chernicoff:2012iq}: for the transversal motion the drag force can increase or decrease, while for the parallel motion the drag force increases in general (except for sufficiently large negative values of $\lgb$). We also plotted the drag force as a function of the quark velocity (Fig. \ref{fig:drag2}). In general, the drag is increased for larger velocities and there is a divergence in the limit $v\to1$, similarly to what occurred in \cite{Chernicoff:2012iq}.

\subsection{Jet quenching parameter} \label{subsec:jet}

Results from RHIC \cite{Arsene:2003yk, Adler:2003ii, Back:2003ns, Adams:2003im} indicate a strong suppression of particles with high transversal momentum $p_\mt{T}$ in Au-Au collisions, but not in d-Au collisions. The explanation for this phenomenon is that in Au-Au collisions the hot and dense quark gluon plasma is produced, and the jets lose energy due to the interaction with this medium before hadronizing. This energy loss effect is called ``jet quenching'' and can give valuable information as regards the properties of the plasma. One important quantity related to jet quenching is the jet quenching parameter $\hat{q}$, which quantify the change of transverse momentum of the parton per unit length when suffering multiple scattering with the medium. The change in transverse momentum is usually referred to as ``momentum broadening''. 

The jet quenching parameter has been calculated at weak coupling for several models (see \cite{Majumder:2010qh} for a review) and has been consistent with data \cite{Burke:2013yra}. However, the assumption of weak coupling is still not well justified, since different energy scales are involved in heavy ion collision experiments and thus a calculation at strongly coupling may be necessary. This motivates a non-perturbative definition of the jet quenching parameter.  The non-perturbative definition of the jet quenching parameter and its first computation using holography was done in \cite{Liu:2006ug, Liu:2006he, D'Eramo:2010ak}. After that, it was extended in several directions.\footnote{There are also some attempts of non-perturbative computations of the jet quenching parameter on the lattice (see, for instance \cite{Panero:2013pla, jq4,jq5}).} See, for instance \cite{jq1,jq2,jq3}. 

The non-perturbative definition of the jet quenching parameter $\hat{q}$ was inspired by its perturbative calculation in the so called dipole approximation \cite{Zakharov:1997uu}
\begin{equation}
 \vev{W^A(\mathcal{C})} \simeq \exp\left[-\frac{L^-\ell^2}{4\sqrt{2}}\hat{q}\right],
\end{equation}
where $W^A(\mathcal{C})$ is a rectangular light-like Wilson loop in the adjoint representation with sizes $L^-$ and $\ell$, with $L^-\gg \ell$. Using the holographic dictionary the jet quenching parameter is given in terms of the on-shell Nambu-Goto action whose string worldsheet boundary coincides with the Wilson loop\footnote{The extra factor of 2 comes from the fact that, for large $N_c$, the Wilson loop in the adjoint representation is roughly speaking the square of the Wilson loop in the fundamental representation.}
\begin{equation}
 \hat{q}=\frac{8\sqrt{2}}{L^{-}\ell^2}S^\mt{on-shell}.
\end{equation}
In the case of pure (isotropic) $\mathcal{N}=4$ SYM, the result obtained was \cite{Liu:2006ug, Liu:2006he, D'Eramo:2010ak}
\begin{equation} \label{eq:qiso}
 \hat{q}_\mt{iso}=\frac{\pi^{3/2}\Gamma(\frac{3}{4})}{\Gamma(\frac{5}{4})}\sqrt{\lambda}T^3.
\end{equation}

Here we compute the jet quenching parameter for the anisotropic background with Gauss-Bonnet term (\ref{eq:metric}). A detailed derivation of the formula we used here is presented in Appendix \ref{app:jet}. The parameters involved are the Gauss-Bonnet coupling $\lgb$, the ratio of the anisotropy parameter to temperature $a/T$ and the angles $(\theta,\varphi)$ associated with the direction of motion of the quark and the direction of the momentum broadening, respectively.\footnote{More precisely, $\theta$ is the angle between the direction of motion of the quark and the anisotropic direction. The direction of the momentum broadening is transversal to the direction of motion of the quark and forms an angle $\varphi$ with the $y$-axis. Note that the same symbols $\theta$ and $\varphi$ were used for other observables, but with different meanings.}

\begin{figure}[H]
 \centering
 \includegraphics[width=0.97\linewidth]{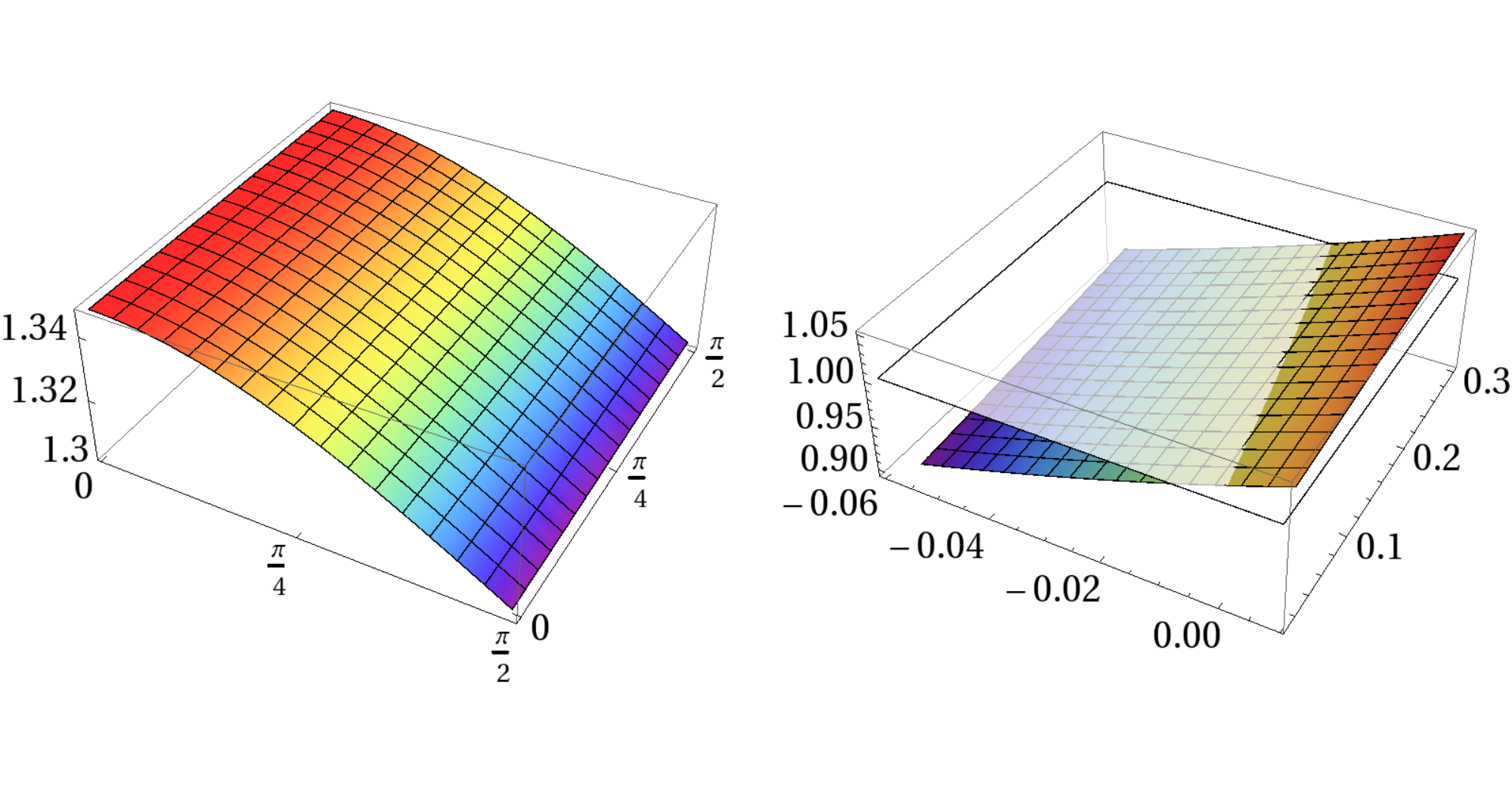}
 \put(-380,50){$\theta$}
 \put(-240,60){$\varphi$}
 \put(-150,40){$\lgb$}
 \put(-17,60){$a/T$}
 \put(-365,200){$\hat{q}/\hat{q}_\mt{iso}$}
 \put(-140,200){$\hat{q}/\hat{q}_\mt{iso}$}
 \caption{Left: Jet quenching parameter as a function of $(\theta, \varphi)$. We have set $\lgb=0.1$ and $a/T=0.33$. Right: The jet quenching parameter as a function of $(\lgb, \frac{a}{T})$. We have set $\theta=\varphi=\pi/4$. Both plots were normalized by the isotropic result (\ref{eq:qiso}).}
 \label{fig:jetplot}
\end{figure}
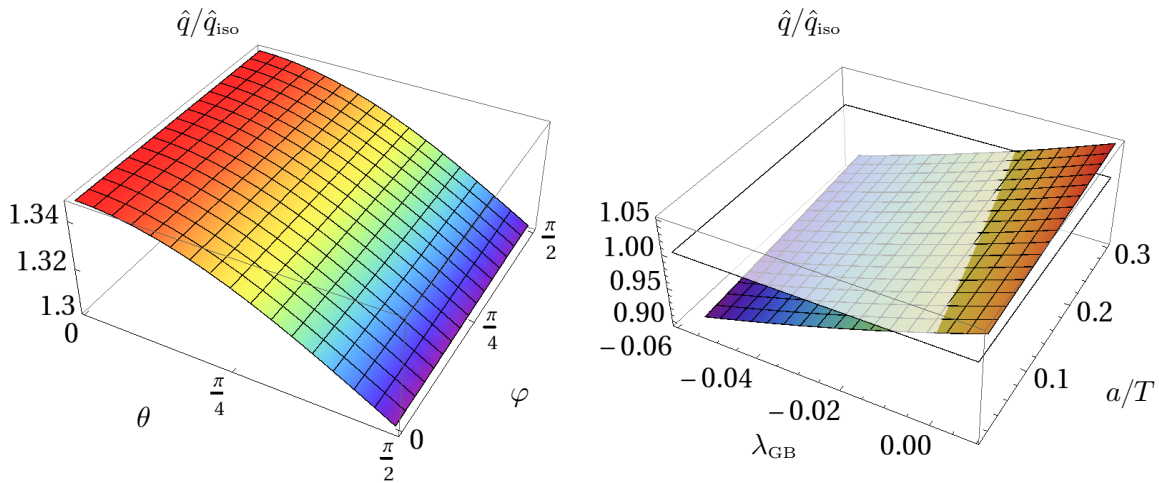

Our results are summarized in Fig. \ref{fig:jetplot}. Similarly to the drag force computation of the previous subsection, the effect of the Gauss-Bonnet coupling is controlled by its sign: the jet quenching parameter is increased for $\lgb>0$ and decreased for $\lgb<0$. The effect of the anisotropy, in the small anisotropy limit, is to increase the jet quenching parameter as it occurred in \cite{Giataganas:2012zy,Chernicoff:2012gu,Giataganas:2013lga}, with the highest increase taking place when the quark moves in the anisotropic direction. We also verified that, for a fixed value of $\theta$, the jet quenching parameter is slightly larger for the momentum broadening taking place in the anisotropic ($\varphi=\pi/2$) direction than in the transversal direction to the anisotropy ($\varphi=0$).
As argued in \cite{Burke:2013yra}, weak coupling models of jet quenching are in general lower than the value obtained at strong coupling for $\mathcal{N}=4$ SYM (\ref{eq:qiso}). If we would expect a smooth interpolation between the weak and strong coupling values, the case $\lgb<0$ would become particularly interesting since it decreases the $\mathcal{N}=4$ SYM result. The same decreasing effect was also found in \cite{Zhang:2012jd}, where one considered fluctuations of the string worldsheet, and in \cite{Armesto:2006zv}, where curvature corrections of type $\alpha'^3R^4$ in the AdS-Schwarzschild background were taken into account. 

\subsection{Quarkonium static potential} \label{subsec:qq}

Quarkonium mesons are produced in the early stages of heavy ion collisions, before the creation of the QGP. As they are much more tightly bound and smaller than ordinary `light' hadrons, they can survive as bound states in the QGP at temperatures above the deconfinement temperature up to some dissociation temperature. This property, together with the fact that their interaction with the thermal medium is comparatively stronger than their interaction with the hadronic matter formed after hadronization, makes the quarkonium mesons excellent probes to study the QGP formed in heavy ion collisions \cite{helmut}.

Here we study the static quarkonium potential in a strongly coupled plasma dual to the gravity theory described in Section \ref{sec:sol}. In particular, we analyze how the anisotropy and the higher derivative terms affect the potential energy and the screening length of a heavy quark-antiquark pair. The holographic studies of this quantity were initiated in \cite{qq1,qq2}, for infinitely heavy quarks in the $\mathcal{N}=4$ SYM theory, and since then several extensions of this work have been performed. See, for instance \cite{Fadafan3,qqDiego,qq3,qq4,qq5,qq6,qq7,qq8,qq9,qq10,qq11,qq12}. Higher derivative corrections to the quarkonium potential were considered in \cite{Fadafan3,qq8} and the effects of anisotropy were taken into account in \cite{Giataganas:2012zy,qqDiego}.

The static quarkonium potential can be extracted from the expectation value of a Wilson loop
\be
\lim_{\mathcal{T} \rightarrow \infty} \left< W(\mathcal{C}) \right> \sim e^{i \mathcal{T} \, (V_{Q\bar{Q}}+2M_Q )},
\label{eq:wilson}
\ee
where $\mathcal{C}$ is a rectangular loop with time extension $\mathcal{T}$ and spatial extension $L$,  $V_{Q\bar{Q}}$  is the quark-antiquark potential energy and $M_Q$ is the quark  mass. The Wilson loop can be viewed as a static quark-antiquark pair separated by a distance $L$. In the gravity side, the pair is described by an open string with both endpoints attached to a D7-brane sitting at some AdS radial position, which determines the quark mass ($M_Q \sim 1/u $). For simplicity, we will work in the case where the D7-brane is at the boundary of AdS and, consequently, the quark is infinitely heavy and non-dynamical.

In the large $N_c$ and large $\lambda$ limits the Wilson loop of Eq. (\ref{eq:wilson}) can be calculated in the gravity side by the expression
\be
\lim_{\mathcal{T} \rightarrow \infty} \left< W(\mathcal{C}) \right> = e^{i S^{( \mt{on-shell} )}},
\ee
where $S^{( \mt{on-shell} )}$ is the on-shell Nambu-Goto action of a U-shaped string whose worldsheet boundary coincides with the curve $\mathcal{C}$. The quarkonium potential is thus obtained as
\be
V_{Q\bar{Q}} =\frac{ S^{( \mt{on-shell} )}}{\mathcal{T}} - 2 M_Q,
\ee
where the quark mass $M_Q$ is determined by evaluating the Nambu-Goto action of a straight string connecting the boundary to the horizon. Given the rotational symmetry in the $xy$-plane, we can assume the quark-antiquark pair to lie in the $xz$-plane, forming an angle $\theta$ with the $z$-direction. Since we want to focus on the results, we leave the details of the calculation of $V_{Q\bar{Q}}(L)$ in Appendix \ref{app:Vqq}. 

\begin{figure}[h!]
    \begin{center}
\Large
        \includegraphics[width=0.66\textwidth]{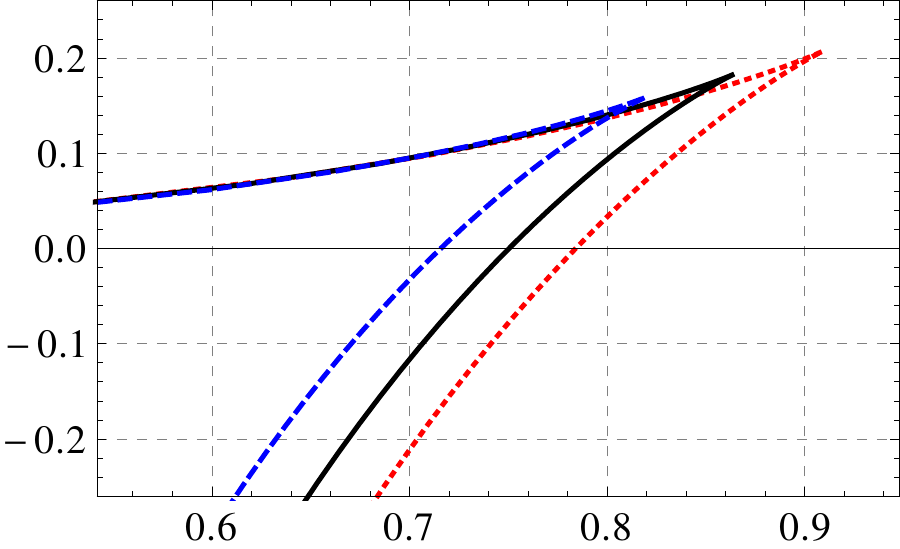}
         \put(-315,90){\rotatebox{90}{$V_{Q\bar{Q}}$}}
         \put(-10,-10){$L$}
        \caption{Quark-antiquark potential $V_{Q\bar{Q}}$ as a function of their separation $L$ for different values of the Gauss-Bonnet coupling: $\lgb = -0.1$ (red, dotted), $\lgb=0$ (black, solid) and $\lgb=0.1$ (blue, dashed). For all curves $a/T \approx 0.3$ and $\theta = \pi/4$.}
        \label{fig:plotVqq}
    \end{center}
\end{figure}

First, let us discuss some general properties of $V_{Q\bar{Q}}(L)$. From Fig. \ref{fig:plotVqq}, we see that $V_{Q\bar{Q}}$ only exists up to a maximum separation length $L_{\text{max}}$. For each value of $L \leq L_{\text{max}}$ there are two possible string configurations corresponding to the upper and lower parts of $V_{Q\bar{Q}}$. It turns out that only the lower part of $V_{Q\bar{Q}}$ represents a physical solution \cite{qq3}. Note that $V_{Q\bar{Q}} = 0$ at some point $L=L_s$, usually referred to as ``screening length''. Since $V_{Q\bar{Q}}$ represents the difference between the energy and mass of the quarkonium, a negative value of the potential ($L \leq L_s$) represents a situation where the U-shaped string (bound state) is energetically favorable over the configuration with two straight strings (unbound state). On the other hand, when the potential is positive ($L \geq L_s$), the opposite happens and the unbound configuration is energetically favorable.\footnote{However, we emphasize that the solution for $V_{Q\bar{Q}}$ is not valid when $L \geq L_s$. In this case the quark-antiquark interaction is completely screened by the presence of QGP between them and, as a consequence, their separation can be increased with no additional energy cost. This implies that the potential is actually constant for $L \geq L_s$. The dual gravity picture can be understood as follows: as we increase the quark-antiquark separation, the U-shaped string connecting the quarks eventually touches the horizon. At this point the string can minimize its energy by splitting into straight strings connecting the boundary of AdS to the horizon.} Another point is that the screening of a plasma is weaker for large $L_s$ and stronger for small $L_s$. This is because $L_s$ represents the separation in which the interaction between the quark and the antiquark becomes completely screened by the medium. 

\begin{figure}[H]
\begin{center}
\begin{tabular}{cc}
\setlength{\unitlength}{1cm}
\hspace{-0.9cm}
\includegraphics[width=7cm]{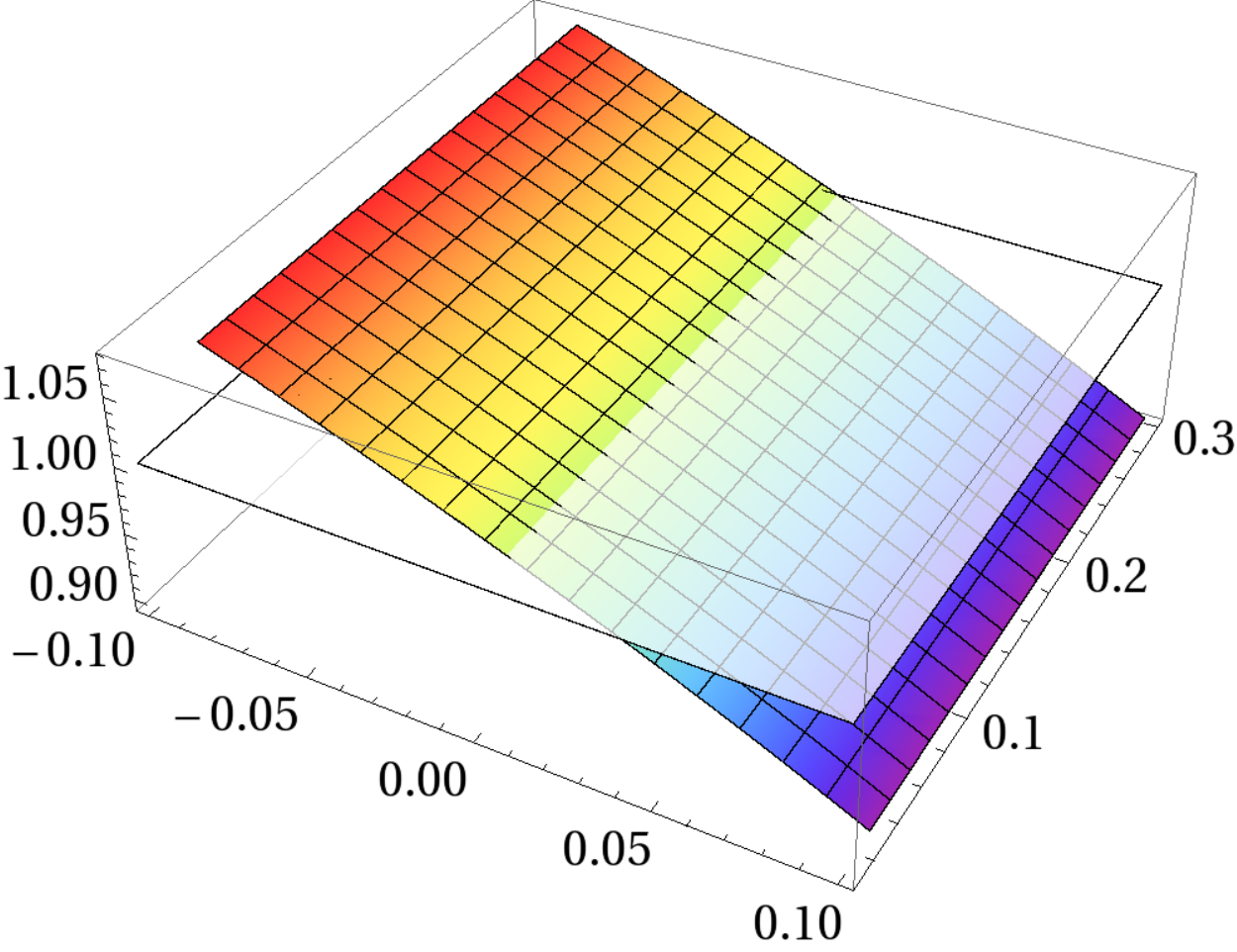} 
\qquad\qquad & 
\includegraphics[width=7cm]{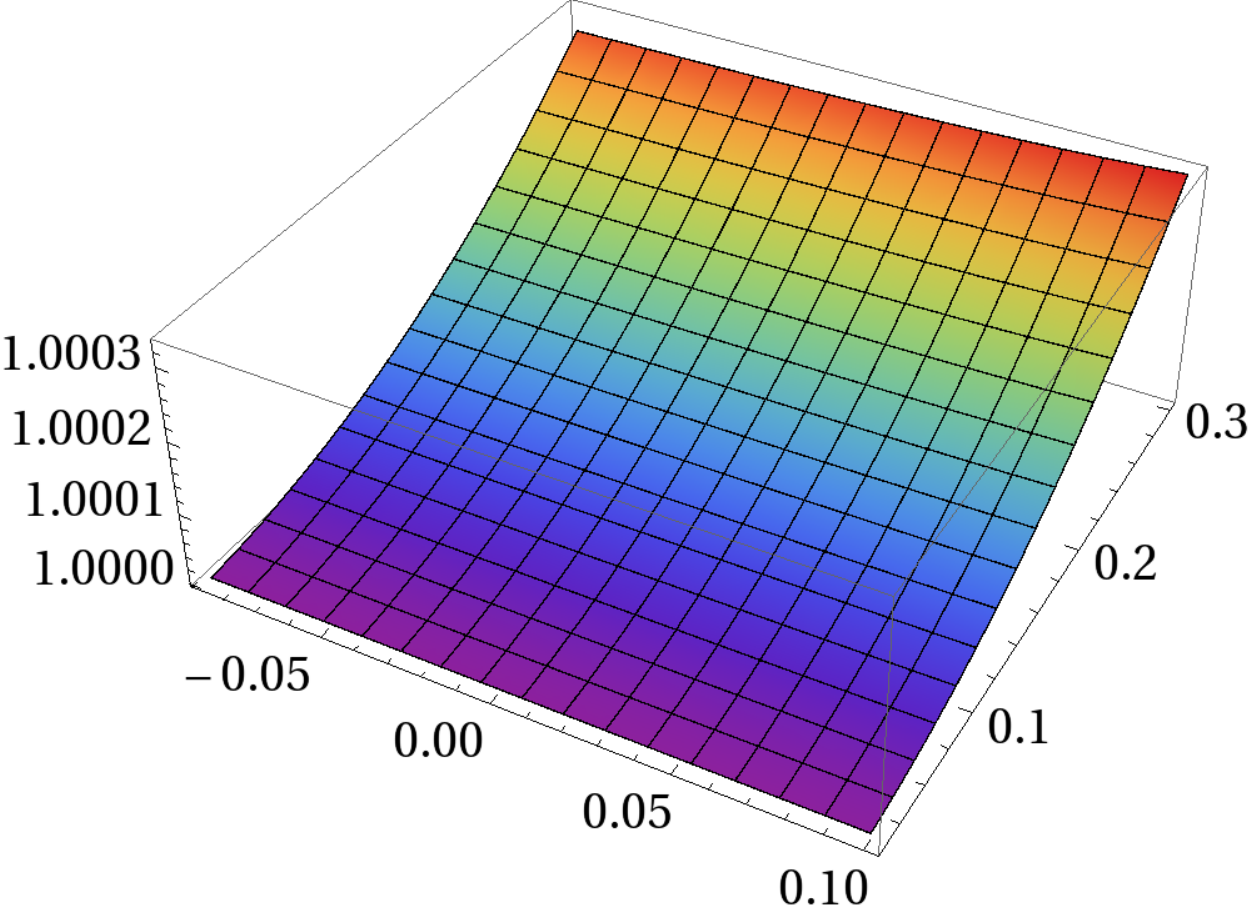}
\qquad
  \put(-385,10){$\lgb$}
         \put(-260,40){$a/T$}
         \put(-145,10){$\lgb$}
         \put(-17,40){$a/T$}
         \put(-365,160){$L_s/L_{\text{iso}}$}
         \put(-120,155){$L_{\perp}/L_{||}$}
\\
(a) & (b)\\
\end{tabular}
\end{center}
\caption{ (a) Screening length $L_s(\lgb,a)$ normalized with respect to the isotropic result $L_{\text{iso}}=L_s(\lgb=0,a=0)$ for $\theta=0$. (b) Ratio $L_{\perp}/L_{||}$, where $L_{\perp}$ is the screening length calculated at $\theta = \pi/2$, and $L_{||}$ is the screening length calculated at $\theta =0$.} 
\label{fig:Ls}
\end{figure}

\begin{figure}[H]
    \begin{center}
\Large
        \includegraphics[width=0.7\textwidth]{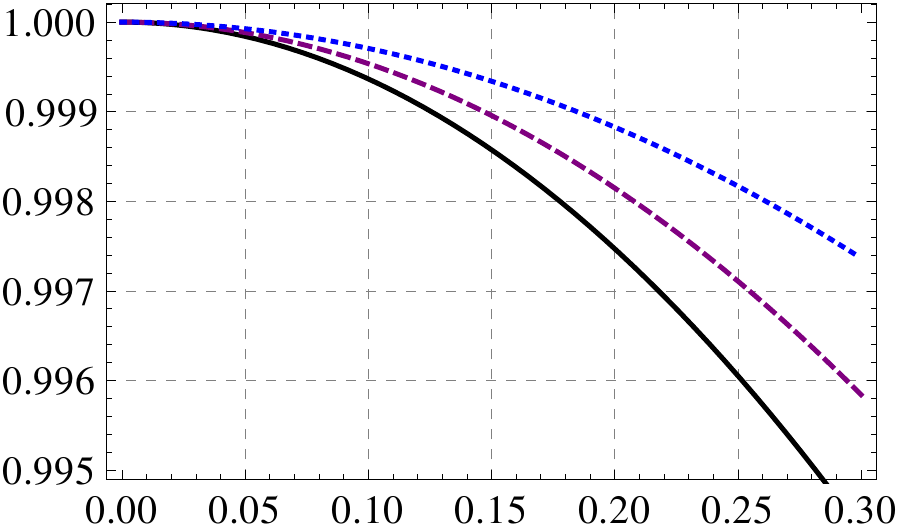}
         \put(-325,90){\rotatebox{90}{$L_s$}}
         \put(-20,-15){$a/T$}
        \caption{Screening length $L_s$ as a function of $a/T$ for three different quarkonium orientations: $\theta=0$ (black, solid), $\theta =\pi/4$ (purple, dashed) and $\theta=\pi/2$ (blue, dotted). The Gauss-Bonnet coupling is fixed $\lgb=0$.}
        \label{fig:plotLa}
    \end{center}
\end{figure}

Fig. \ref{fig:plotVqq} shows that positive values of $\lgb$ decrease the screening length, while negative values of $\lgb$ increase this quantity. This effect can be better visualized in Fig. \ref{fig:Ls} (a), where the screening length is presented as a function of $(\lgb,a)$. Now let us discuss the effect of the anisotropy. First of all, Fig. \ref{fig:Ls} (b) shows that the screening length for a quarkonium oriented along the anisotropic direction ($\theta =0$) is always smaller than the corresponding quantity for a quarkonium oriented in the transverse plane ($\theta= \pi/2$). Second, the 2D plot of Fig. \ref{fig:plotLa} reveals that the screening length always decrease as we increase $a/T$, for any orientation of the pair, at fixed $\lgb$. These anisotropic effects are also observed in holographic calculations at strong coupling when the anisotropy is introduced by a magnetic field \cite{qq12} and at weak coupling in calculations based on ``hard-thermal-loop'' resummed perturbative QCD \cite{qq13}. The limit $\lgb \rightarrow 0$ of the above results agrees with the calculations of \cite{Giataganas:2012zy}. We also checked that the limit $a \rightarrow 0$ for $V_{Q\bar{Q}}$ agrees with the results of \cite{Fadafan3} when the quasi-topological coupling constant is zero.\footnote{In the comparison of our results with \cite{Fadafan3}, one should note that the potential of \cite{Fadafan3} is normalized with $1/(\pi \alpha')$, while our results are normalized with $1/(2 \pi \alpha')$.}

\subsection{Photon production} \label{subsec:pp}
The limited extension of the QGP created in heavy ion collisions and the weakness of the electromagnetic interactions imply that this medium should be optically thin. Therefore, the photons produced in the plasma escape from it without subsequent interactions, providing an excellent probe of the conditions of the medium. The holographic studies of this quantity were initiated in \cite{pp1} and extended in several directions, see, for instance \cite{ pp5,pp15,pp2, pp3, pp4, pp6, pp7, pp8, pp9, pp10, pp11, pp12, pp13, pp14, pp16, pp17,pp18,pp19,pp20}. In this section we study how the anisotropy and higher derivative corrections affect the photon production rate in the model described in Section \ref{sec:sol}.

\begin{figure}
    \begin{center}
        \includegraphics[width=0.65\textwidth]{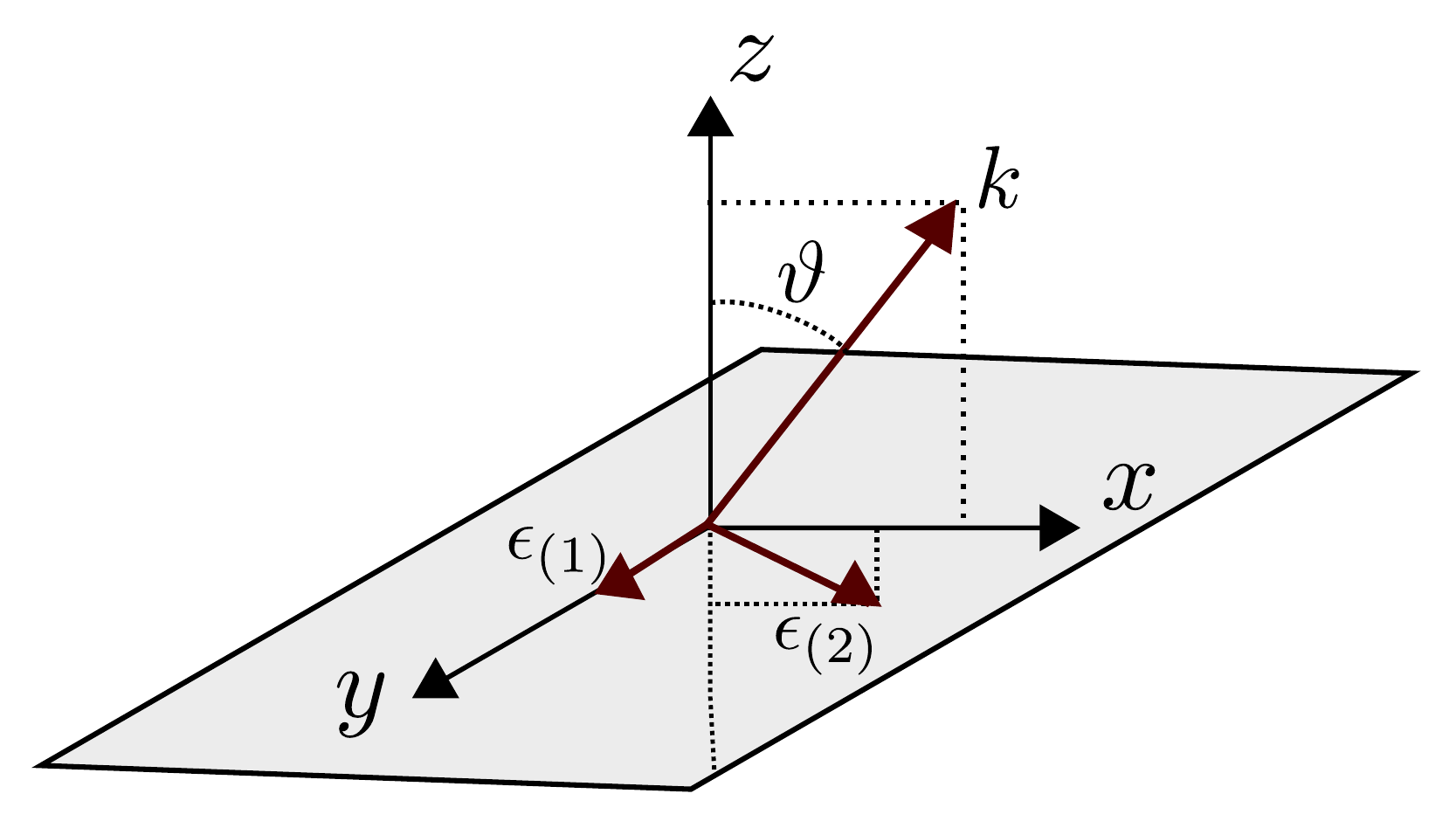}
        \caption{Momentum $\vec k$ and polarization vectors $\vec \epsilon_{(1)}$ and $\vec \epsilon_{(2)}$. The $SO(2)$ rotational symmetry in the $xy$-plane allows us to choose the momentum lying in the $xz$-plane, forming an angle $\vartheta$ with the $z$-direction. Both polarization vectors are orthogonal to $\vec k$. We chose $\vec \epsilon_{(1)}$ oriented along the $y$-direction and $\vec\epsilon_{(2)}$ contained in the $xz$-plane.}
        \label{momentum}
    \end{center}
\end{figure}

Let $\mathcal{L}_0$ be the Lagrangian of the field theory dual to the gravity theory described by the action (\ref{eq:action}). The photon production rate is calculated by adding a dynamical photon to $\mathcal{L}_0$ coupled to the electric charged matter fields, that is,
\be
\mathcal{L} = \mathcal{L}_0+e \jem_{\mu} A^{\mu} -\frac{1}{4}F_{\mu \nu}F^{\mu \nu},
\ee
where $F_{\mu \nu} = \partial_{\mu} A_{\nu}-\partial_{\nu} A_{\mu}$ is the field strength associated to the photon field $A^{\mu}$, $e$ is the electromagnetic coupling constant and $\jem_{\mu}$ is the electromagnetic current. At leading order in $e$, the number of photons emitted per unit time and unit volume is given by \cite{lebellac}
\bea
\frac{d\Gamma_\gamma}{d^3 k} = \frac{e^2}{(2\pi)^3 2|\vec k|}\Phi(k) \,  \eta^{\mu \nu} \chi_{\mu \nu}(k)\Big|_{k^0=|\vec k|}\,,
\label{difftr}
\eea
where $\eta^{\mu \nu}=\text{diag}(-+++)$ is the Minkowski metric, $k^\mu=(k^0,\vec k)$ is the photon null momentum, $\Phi(k)$ is the distribution function and $\chi_{\mu \nu}$ is the spectral density. Assuming thermal equilibrium, the distribution function reduces to the Bose-Einstein distribution $n_B(k^0)=1/(e^{k^0/T}-1)$. The spectral density can be obtained as 
\be
\chi_{\mu\nu}(k)=-2 \mbox{ Im } G^\mt{R}_{\mu \nu}(k),
\label{eqchi}
\ee
where
\bea
G^\mt{R}_{\mu\nu}(k) = -i \int d^4x \, e^{-i k\cdot x}\, \Theta(t) \vev{[\jem_\mu(x),\jem_\nu(0)]}
\eea
is the retarded correlator of two electromagnetic currents $\jem_\mu$ and the above expectation value is taken in the thermal equilibrium state. The Ward identity $k^\mu\chi_{\mu\nu}=0$ for null $k^\mu$ implies that only the transverse spectral functions contribute in the calculation of the trace of the spectral density, that is,
\be
\eta^{\mu \nu} \chi_{\mu \nu} =\sum_{s=1,2} \epsilon^\mu_{(s)}(\vec k)\,  \epsilon^\nu_{(s)}(\vec k)\, \chi_{\mu\nu}(k)\Big|_{k^0=|\vec k|}\,.
\ee
Using the above formula, the differential photon production rate can be rewritten as
\bea
\frac{d\Gamma_\gamma}{d^3 k} = \frac{e^2}{(2\pi)^3 2|\vec k|}\Phi(k)\sum_{s=1,2} \epsilon^\mu_{(s)}(\vec k)\,  \epsilon^\nu_{(s)}(\vec k)\, \chi_{\mu\nu}(k)\Big|_{k^0=|\vec k|}\,,
\label{diff}
\eea
where $\epsilon^\mu_{(1)}$ and $\epsilon^\mu_{(2)}$  are mutually orthogonal polarization vectors that are also orthogonal to $k^{\mu}$. By the $SO(2)$ symmetry in the $xy$-plane of our model we can choose $\vec{k}$ to lie in the $xz$-plane -- see Fig.~\ref{momentum}. Following \cite{pp5,pp15}, we set
\be
\vec k = k_0 (\sin \vartheta, 0, \cos \vartheta) \,.
\ee
With this choice of $\vec{k}$ the polarization vectors can be chosen as
\be
\vec\epsilon_{(1)}=(0,1,0)\sac \vec\epsilon_{(2)} = (\cos \vartheta, 0, - \sin \vartheta) \,.
\ee
For later purposes we split the trace of the spectral density into two parts
\be
\eta^{\mu \nu} \chi_{\mu \nu} = \chi_{(1)}+\chi_{(2)},
\ee
where $\chi_{(s)}$ is proportional to the number of photons emitted with polarization $\vec{\epsilon}_{(s)}$. These quantities are given by
\begin{align}
\chi_{(1)} & =  \epsilon^\mu_{(1)}  \epsilon^\nu_{(1)} \chi_{\mu \nu} = \chi_{yy} \nonumber \\ 
\chi_{(2)} & = \epsilon^\mu_{(2)}\,  \epsilon^\nu_{(2)}\, \chi_{\mu\nu} = 
\cos^2 \vartheta \, \chi_{xx} + \sin^2 \vartheta \, \chi_{zz} 
- 2 \cos \vartheta \sin \vartheta  \, \chi_{xz} \,. 
\label{eqchi2}
\end{align}
We now proceed to explain how to compute the retarded Green function of two electromagnetic currents using holography. It turns out that the gravity theory dual to the field theory described by the Lagrangian $\mathcal{L}$ is simply obtained by adding a $U(1)$ kinetic term to the action (\ref{eq:action}). As we are dealing with a bottom-up model, we consider a five-dimensional $U(1)$ kinetic term of the form,
\be
S_{U(1)} = -K \int d^5 x \, F_{mn} F^{mn},
\label{actionU1}
\ee
where $F_{mn} = \partial_m A_n - \partial_n A_m$ is the field strength associated to the gauge field $A_m$ $(m=0,1, 2, 3, 4)$ and $K$ is a constant.\footnote{In top-down calculations, $K$ is proportional to the number of electrically charged degrees of freedom times the number of colors in the dual gauge theory. For instance, when photons are produced from adjoint matter we have $K \propto N_c^2$ \cite{pp1}, while for fundamental fields, $K \propto N_c N_f$ \cite{pp3, pp5, pp15,pp19}. In bottom-up models, this constant can be chosen freely and, since we are only interested in ratios of spectral densities (which are proportional to $K$), this constant will play no role in our analysis.} Let $A_{\mu}$ ($\mu=0,1,2,3$) denote the components of this gauge field along the gauge theory coordinates $(t,\vec{x})$ and $A_4 = A_u$ denote the component along the radial coordinate of $AdS$. In order to simplify our calculations, we gauge fix $A_u = 0$. Our final results, however, will be written only in terms of gauge invariant quantities, in such a way that this gauge choice will not be relevant.

Given the translation invariance of our model, we can Fourier decompose the gauge field $A_{\mu}$ as
\be
A_{\mu}(t,\vec{x},u) = \int \frac{d^4k}{(2 \pi)^4} e^{-i k^0 t+i \vec{k} \cdot \vec{x}} A_{\mu} (k^0,\vec{k},u)\,.
\ee
The equations of motion derived from (\ref{actionU1}) are given by
\be
\partial_{\mu} \big( \sqrt{-g} g^ {\mu \alpha} g^{\nu \beta} F_{\alpha \beta} \big) =0\,.
\ee
In terms of the gauge invariant quantities $E_i = \partial_0 A_i - \partial_y A_i$, the above equations of motion split into a decoupled equation for $E_y$,
\be
E_y''+\big( \text{log}\, \sqrt{-g} g^{uu} g^{yy} \big)' E_y'-\frac{\overline{k}^2}{g^{uu}} E_y=0\, ,
\ee
and a system of two coupled equations for $E_x$ and $E_z$,\footnote{In the derivation of the equations of motion for $E_x$ and $E_z$ we used the constraint $g^{\alpha \beta} k_{\alpha} A_{\beta}'=0$.}
\begin{align}
& E_{x}''+\left[\left(\mathrm{log} \sqrt{-g} g^{uu}g^{xx}\right)'-\left(\mathrm{log}\frac{g^{xx}}{g^{tt}}\right)'\frac{k_{x}^{2}}{\overline{k}^{2}}g^{xx}\right]E_{x}'-\frac{\overline{k}^{2}}{g^{uu}}E_{x}-\left(\mathrm{log}\frac{g^{xx}}{g^{tt}}\right)'\frac{k_{z}k_{x}}{\overline{k}^{2}}g^{zz}E_{z}'=0\,, \nonumber \\
& E_{z}''+\left[\left(\mathrm{log} \sqrt{-g} g^{uu}g^{zz}\right)'-\left(\mathrm{log}\frac{g^{zz}}{g^{tt}}\right)'\frac{k_{z}^{2}}{\overline{k}^{2}}g^{zz}\right]E_{z}'-\frac{\overline{k}^{2}}{g^{uu}}E_{z}-\left(\mathrm{log}\frac{g^{zz}}{g^{tt}}\right)'\frac{k_{z}k_{x}}{\overline{k}^{2}}g^{xx}E_{x}'=0\,, \label{eq:ExEz}
\end{align}
where the primes denote derivatives with respect to $u$ and $\overline{k}^2 \equiv g^{\alpha \beta} k_{\alpha} k_{\beta}$. Note that the above equations are written in momentum space.

The action (\ref{actionU1}) can be written in terms of the gauge invariant fields $E_i$ as
\begin{multline}
S_{\epsilon}=-2K\int dt\,d\vec{x}\,\frac{\sqrt{-g}g^{uu}}{k_{0}^{2}\overline{k}^{2}}\left[\left(-g^{tt}k_{0}^{2}-g^{zz}k_{z}^{2}\right)g^{xx}E_{x}E_{x}'-\overline{k}^{2}g^{yy}E_{y}E_{y}'+\right.\\
+g^{xx}g^{zz}k_{x}k_{z}\left(E_{x}E_{z}\right)'+\left(-g^{tt}k_{0}^{2}-g^{xx}k_{x}^{2}\right)g^{zz}E_{z}E_{z}'\Biggr]_{u=\epsilon}\label{eq:boundarycampos}\,.
\end{multline}
The retarded correlators are obtained by taking functional derivatives of the above action with respect to the boundary values of the gauge fields $A^{\mu (0)}$. In the computation of $\chi_{(1)}$ and $\chi_{(2)}$ we only need the spatial correlators $G^{\text{R}}_{xx}$, $G^{\text{R}}_{yy}$, $G^{\text{R}}_{zz}$, and $G^{\text{R}}_{xz} = G^{\text{R}}_{zx}$. This correlators can be obtained in terms of the $E_i$'s as
\be
G^{\text{R}}_{i j}=\frac{ \delta^2 S_{\epsilon}}{\delta A^{i (0)} \delta A^{j (0)}} = k_0^2 \frac{ \delta^2 S_{\epsilon}}{\delta E_i^{(0)} \delta E_j^{(0)}} \,,
\label{eqGii}
\ee
where $E_i^{(0)}$ is the boundary value of the gauge field $E_i$.

As the mode $E_y$ does not couple to the other modes, the spectral density for photons with polarization $\vec{\epsilon}_{(1)}$ can be obtained by applying the prescription of \cite{recipe}. The retarded correlator reads
\be
G^{\text{R}}_{yy} = k_0^2 \frac{\delta^2 S_{\epsilon}}{\delta E_y^{(0) 2}} =-\frac{4 K}{k_0^2} \sqrt{-g} g^{uu} g^{yy} \frac{E_y'(k,u)}{E_y(k,u)} \Big|_{u \rightarrow 0}.
\ee
The corresponding spectral density is then given by
\be
\chi_{(1)} = \chi_{yy} =-2 \text{Im}\, G^{\text{R}}_{yy} =\frac{8 K}{k_0^2} \text{Im}\, \Big[ \sqrt{-g} g^{uu} g^{yy} \frac{E_y'(k,u)}{E_y(k,u)} \Big]_{u \rightarrow 0}.
\ee

The computation of $\chi_{(2)}$ is more involved, because of the coupling between $E_x$ and $E_z$. We face this problem by following the technique developed in \cite{pp5} to deal with mixed operators. First, we write a near-boundary expression for the fields $E_x$ and $E_z$,
\begin{align}
 E_x=E_x^{(0)}+u^2E_x^{(2)}\cos\vartheta+u^4E_x^{(4)}+O(u^6)\,,\nonumber\\
 E_z=E_z^{(0)}-u^2E_x^{(2)}\sin\vartheta+u^4E_z^{(4)}+O(u^6)\,.
\end{align}
The form of the second order coefficients was chosen such that the equations of motion (\ref{eq:ExEz}) are satisfied. The equations of motion also determine the coefficients $E_x^{(4)}$ and $E_z^{(4)}$ in terms of the lower order coefficients,
\begin{align}
E_x^{(4)} & = \frac{a^2 \lgb  \cos \vartheta}{96 (1-B_0) (1-4 \lgb)} \left(3 k_0^2 (B_0-2 \lgb ) E_x^{(0)} \cos \vartheta+8 (1-2 B_0) E_x^{(2)}\right) \,, \nonumber\\
E_z^{(4)} & =\frac{a^2}{192 \sqrt{1-4 \lgb }} \left[3 k_0^2 (\lgb -B_0)  \left(E_x^{(0)} \sin \vartheta - E_z^{(0)}  \cos \vartheta \right) \cos \vartheta -8  B_0 E_x^{(2)} \sin \vartheta\right]\,.
\end{align}
The remaining coefficients $E_x^{(0)}$, $E_z^{(0)}$ and $E_x^{(2)}$ can be extracted from the numerical solution. With the above expressions the boundary action (\ref{eq:boundarycampos}) takes the form
\begin{equation}
S_{\epsilon} =\sqrt{B_0}K\left[-\frac{1}{2}\left(E_x^{(0)}\sin\vartheta+E_z^{(0)}\cos\vartheta \right)^2 -\frac{4}{B_0 k_0^2}\left( E_x^{(0)}E_x^{(2)}\cos\vartheta+ E_z^{(0)}E_x^{(2)}\sin\vartheta\right)\right].
\end{equation}
Finally, using (\ref{eqchi}), (\ref{eqchi2}), and (\ref{eqGii}) we can show that
\be
\chi_{(2)} = \frac{16 K}{ \sqrt{B_0}} \,\text{Im}\, \left[\frac{\delta E_x^{(2)}}{\delta E_x^{(0)}} \, \text{cos}\, \vartheta-\frac{\delta E_x^{(2)}}{\delta E_z^{(0)}} \, \text{sin}\, \vartheta  \right],
\ee
where the functional derivatives $\delta E_x^{(2)} / \delta E_x^{(0)}$ and $\delta E_x^{(2)} / \delta E_z^{(0)}$ are calculated according to the prescription given in \cite{pp5}.


The trace of the spectral density $\chi_{\,\,\mu}^{\mu}=\chi_{(1)}+\chi_{(2)}$ is a function of the parameters $(\lgb, a, \vartheta, \uh,k^0)$. In order to study the effects of the anisotropy parameter and the Gauss-Bonnet coupling, we computed $\chi_{\,\,\mu}^{\mu}$ for several values of $(\lgb,a,\vartheta)$, choosing as normalization the isotropic result
\begin{equation} \label{eq:ppiso}
 \chi_{\mt{iso}}=\chi_{\,\,\mu}^{\mu}(\lgb=0, a=0).
\end{equation}

\begin{figure}[H]
\begin{center}
\begin{tabular}{cc}
\setlength{\unitlength}{1cm}
\hspace{-0.9cm}
\includegraphics[width=7cm]{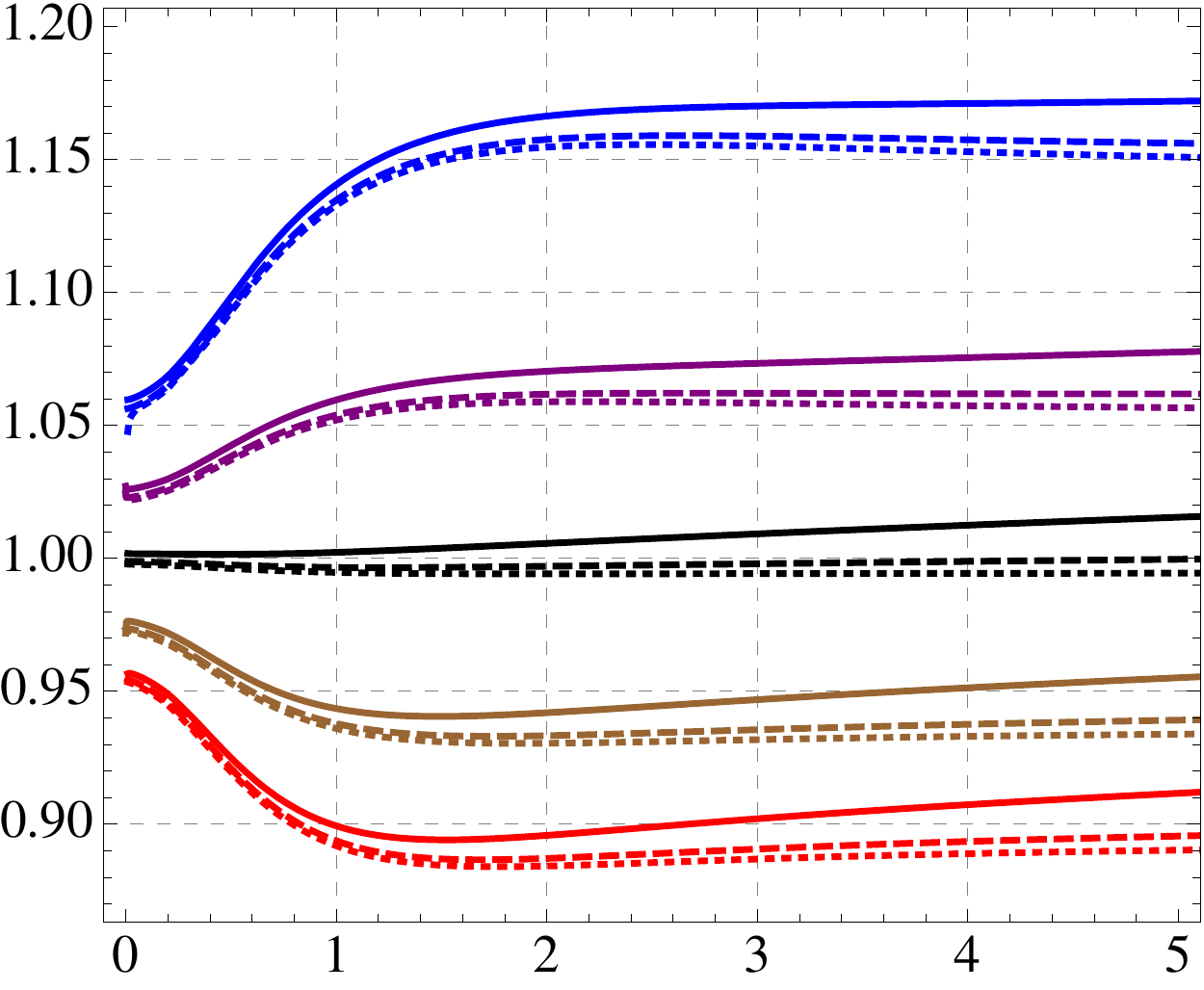} 
\qquad\qquad & 
\includegraphics[width=7cm]{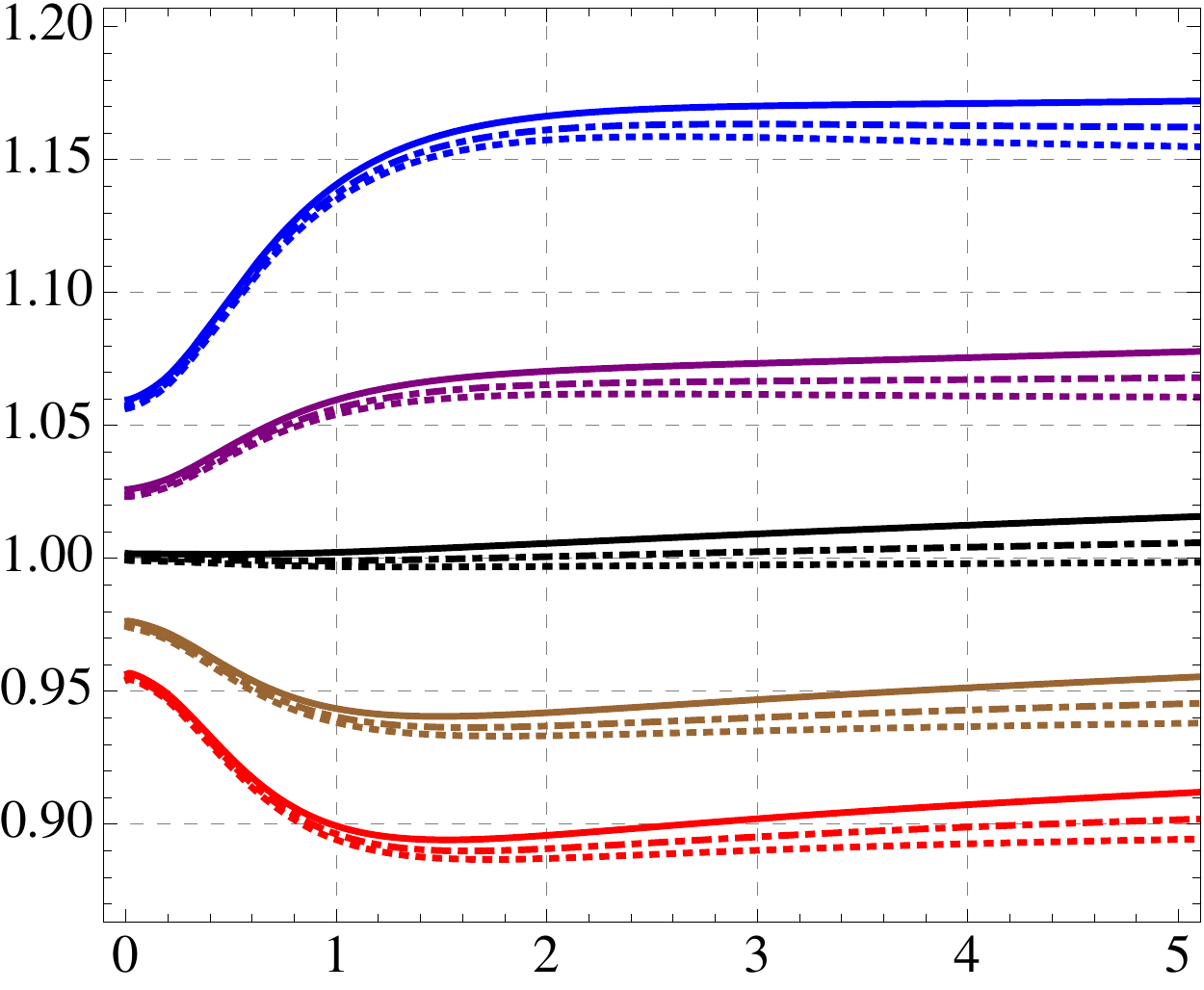}
\qquad
  \put(-455,40){\rotatebox{90}{$\chi_{\,\,\mu}^{\mu}(\lgb, a,\vartheta)/\chi_{\mt{iso}}$}}
         \put(-250,-10){$\wn$}
         \put(-215,40){\rotatebox{90}{$\chi_{\,\,\mu}^{\mu}(\lgb, a,\vartheta)/\chi_{\mt{iso}}$}}
         \put(-17,-10){$\wn$}
\\
(a) & (b)\\

\end{tabular}
\end{center}
\caption{\small The trace of the spectral density $\chi_{\,\,\mu}^{\mu}(\lgb,a,\vartheta)$ normalized with respect to the isotropic result (\ref{eq:ppiso}). All the spectral densities were calculated at the same temperature $T_0=0.316698$. The colors of the curves identify the value of the $\lgb$ parameter as: red curves ($\lgb=-0.1$), brown curves ($\lgb=-0.05$), black curves ($\lgb=0$),  purple curves ($\lgb =0.05$) and blue curves ($\lgb=0.1$). In (a), the angle of emission is fixed ($\vartheta =0$) and we have solid curves $(a=0.2)$, dashed curves $(a=0.1)$ and dotted curves $(a=0)$. In (b), the anisotropy is fixed ($a=0.2$) and we have solid curves $(\vartheta=0)$, dot-dashed curves $(\vartheta = \pi/4)$, and dotted curves $(\vartheta = \pi/2)$.}
\label{plotschi}
\end{figure}

Our comparison with the isotropic result was made at fixed temperature  $T_0=0.32$.\footnote{Doing this, one must note that the temperature $T$ of the system is a function of $(\lgb,a,\uh)$ and, consequently, it changes as we vary these parameters. Therefore, we need to adjust $\uh$ in such a way that all the spectral densities are calculated at same temperature $T_0$, defined by $T_0 = T(\lgb=0,a=0,\uh=1)$.} The results for the ratio $\chi_{\,\,\mu}^{\mu}/\chi_{\mt{iso}}$ as a function of the dimensionless frequency $\wn = k^0/ 2 \pi T_0$ are presented in Fig. \ref{plotschi}. For an anisotropic plasma, we have $\chi_{(1)} \neq \chi_{(2)}$. However, in our case  the smallness of the anisotropy parameter $a$ makes these two quantities almost equal, presenting a very similar behavior as a function of $\wn$, so we chose to plot only the total spectral density instead of plotting the two spectral densities separately. At least, we observed that $\chi_{(1)}$ is slightly bigger than $\chi_{(2)}$, as was the case in \cite{pp5,pp15}.
We also verified that our results reproduce the calculations of \cite{pp18} in the limit $a \rightarrow 0$ and that they are consistent with anisotropic calculations of \cite{pp5}  in the limit $\lgb \rightarrow 0$ and small values of $a/T$.

From Fig. \ref{plotschi} it is clear that the effect of the Gauss-Bonnet coupling is to increase or decrease the photon production rate, depending on whether $\lgb >0$ or $\lgb<0$, respectively. 
The main effect of the anisotropy parameter is to increase the photon production rate. At small frequencies, $\chi_{\,\,\mu}^{\mu}$ does not depend strongly on $a$. For generic frequencies, the $\chi_{\,\,\mu}^{\mu}$ is higher for photons with longitudinal wave vectors $(\vartheta=0)$ than for the ones with transverse wave vectors $(\vartheta=\pi/2)$. One qualitative difference between the corrections introduced by $\lgb$ and $a$ is their dependence on the frequency. Looking at the curves for $a=0$ in Fig. \ref{plotschi}, we see that the Gauss-Bonnet correction reaches a constant value after a sufficiently large value of $\wn$. On the other hand, the effect of the anisotropy parameter $a$ is enhanced as we increase $\wn$.

\begin{figure}[H]
    \begin{center}
\Large
        \includegraphics[width=0.67\textwidth]{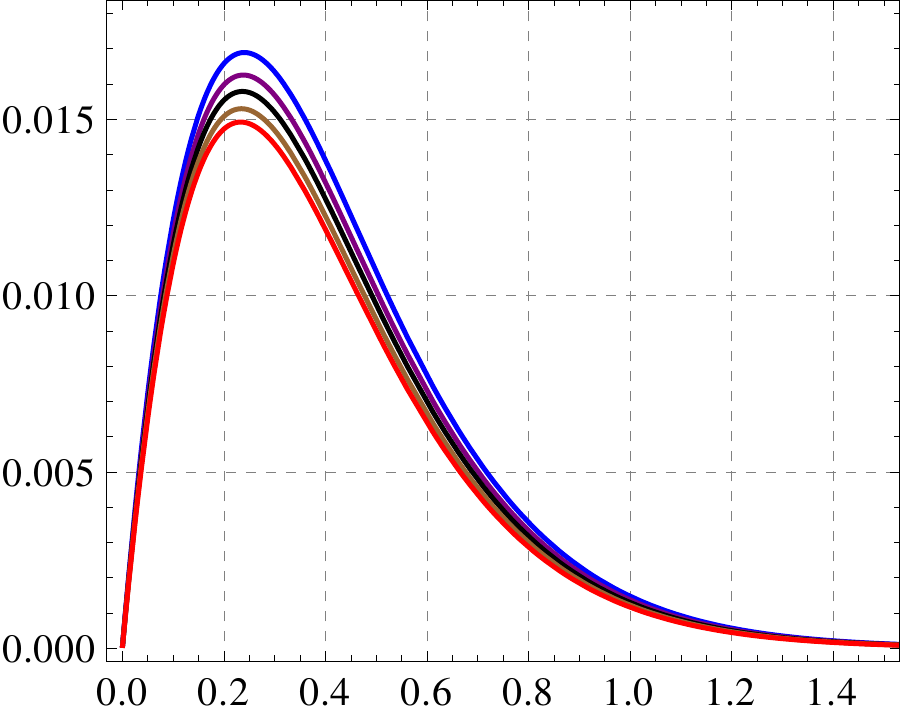}
         \put(-330,90){\rotatebox{90}{ $\frac{-1}{ 4 K e^2 T_0^3}\frac{d\Gamma_\gamma}{d\cos\vartheta\, dk^0}$  }}
         \put(-20,-15){$\wn$}
        \caption{Total photon production rate as a function of $\wn = k^0/ 2\pi T_0$. From top to bottom, the value of the Gauss-Bonnet coupling is identified as $\lgb=0.1$ (blue),  $\lgb=0.05$ (purple), $\lgb=0$ (black), $\lgb=-0.05$ (brown), $\lgb=-0.1$ (red). We have fixed $\vartheta = 0$ and $a=0.2$. The results for different angles are very similar to the plot above due to the smallness of the anisotropy.}
        \label{fig:TPP}
    \end{center}
\end{figure}

It is also interesting to analyze how the anisotropy and the Gauss-Bonnet term affects the total photon production (\ref{difftr}), which can be expressed as
\be
\frac{-1}{ 4 K e^2 T_0^3}\frac{d\Gamma_\gamma}{d\cos\vartheta\, dk^0} =\frac{\wn}{ 32 K \pi^3 T_0^2} \frac{1}{e^{2\pi \wn}-1} \big(\chi_{(1)}+\chi_{(2)} \big)
\ee
This quantity is shown in Fig. \ref{fig:TPP}, for different values of $\lgb$ and $\vartheta$. From Fig. \ref{fig:TPP} we see that, for $\lgb > 0$, the peak in the spectrum of photons becomes higher, widens and gets shifted to the right. For $\lgb < 0$, the peak becomes smaller, sharpens and gets shifted to the left. This should be contrasted with the results of \cite{pp12} for a top-down higher derivative correction of the form $\alpha'^3 R^4$, where the peak in the spectrum becomes higher, sharpens and gets shifted to the left, approaching the weak coupling result \cite{pp1}, which shows a very sharp peak at small $\wn$ in the photon spectrum. Therefore, the inclusion of the $\alpha'^3 R^4$ correction (which corresponds to a finite 't Hooft coupling correction in the gauge theory) goes into the direction of the weak coupling results, while this does not seems to be possible in the case of Gauss-Bonnet. However, a partial agreement between these two types of corrections is found when $\lgb<0$, where the peak in the photon spectrum sharpens and moves to the left, but it also becomes smaller, contrary to what happens at weak coupling. We can understand this partial agreement  noting that, for $\lgb <0$, the ratio $\eta/s = (1-4\lgb)/(4 \pi)$ increases, which also happens with $\eta/s$ when finite 't Hooft coupling corrections were taking into account. Since at weak coupling the shear viscosity over the entropy density ratio is proportional to the mean free path of momentum isotropization, we can associate the approaching of the weak coupling results (negative $\lgb$ corrections or $\alpha'$-corrections) with a larger mean free path in both cases.

\section{Discussion} \label{sec:discussion}

We have studied how the anisotropy and higher curvature terms affect several observables relevant to the study of the QGP, namely, the drag force, the jet quenching parameter, the quarkonium static potential and the photon production rate. In the gravity side, the anisotropy was introduced by an external source (an axion linear in the beam direction) that keeps the system in an equilibrium anisotropic state, while the higher curvature correction was chosen to be the Gauss-Bonnet term. 

The effect of the Gauss-Bonnet term in our results are consistent with previous results \cite{Fadafan1, Fadafan2, Fadafan3, pp18} and they are summarized in Table \ref{tab:summary}, where we specify if the value of the observable increases or decreases compared to the case of isotropic $\mathcal{N}=4$ SYM. In this table we also present the result for the shear viscosity over entropy density obtained previously \cite{Jahnke:2014vwa} and the finite 't Hooft corrections of type $\alpha'^3 R^4$ for these observables \cite{Fadafan5, Fadafan6, df15, pp12}.  

\begin{table}[H]
\caption{Summary of the effect of the Gauss-Bonnet coupling $\lgb$ on several observables. We also present the finite 't Hooft corrections of type $\alpha'^3 R^4$  \cite{Fadafan5, Fadafan6, df15, pp12}. The comparison is taken w.r.t. the respective $\mathcal{N}=4$ SYM result at same temperature.}
\vspace{4pt}
 \begin{tabular}{ |c|c|c|c|c|c| } \hline
 &  $\eta/s$ &  Drag force & Jet quenching & Screening length & Photon production \\ \hline\hline
 $\lgb > 0$       & decrease  &   increase   & increase    & decrease &  increase\\ 
 $\lgb < 0$        & increase  &   decrease   & decrease    & increase &  decrease\\ 
 $\alpha'^3 R^4$  & increase  &   increase   & decrease    & decrease &  increase\\ \hline
\end{tabular}
\label{tab:summary}
\end{table}

A possible heuristic interpretation of the increasing/decreasing in the above observables is to correlate these results with the changes in the ratio $\eta/s$. 
At weak coupling, $\eta/s$ is proportional to the mean free path of momentum isotropization of the plasma ($\eta/s \sim \ell_{\text{mfp}}$). Imagining a situation where the mean free path is decreasing, we should expect an external probe to interact more with the medium, increasing the energy loss of the probe and its probability to suffer scattering. As a result, we would obtain an increase in the drag force and the jet quenching parameter. Moreover, a low mean free path would break the connection between a quark-antiquark pair more easily, resulting in a low value of screening length. Finally, a low mean free path would raise the number of collisions per time and, consequently, the number of photons produced in these interactions would increase. Note that this situation matches exactly the case of $\lgb>0$. Of course, the opposite idea applies for $\lgb<0$. Although this reasoning seems to be consistent for the Gauss-Bonnet, it does not work when applied to the $\alpha'^3 R^4$ correction. 

The effect of the anisotropy is similar to what was found previously \cite{Chernicoff:2012iq,Giataganas:2012zy,Chernicoff:2012gu,  qqDiego,pp5,Giataganas:2013lga}. The photon production rate and the quarkonium dissociation length in an anisotropic plasma are bigger than the corresponding quantities in an isotropic plasma at the same temperature. The drag force and the jet quenching parameter in an anisotropic plasma can be bigger or smaller than its isotropic counterparts, depending on several parameters like the quark velocity, the direction of the quark motion, and the direction of momentum broadening. Below we also summarize the effects of the anisotropy with a comparison between the value of the observables along the anisotropic direction ($||$) and along the transverse plane ($\perp$):
\begin{itemize}
 \item Shear viscosity: \quad $\eta_{\perp} > \eta_{||},$
 \item Drag force: \quad $F_\mt{drag}^{\perp} < F_\mt{drag}^{||},$
 \item Jet quenching parameter: \quad $\hat{q}_{\perp} < \hat{q}_{||},$
 \item Screening length: \quad $L_{\perp} > L_{||},$
 \item Photon production rate:\quad $\chi^{\,\,\mu}_{\mu\,\, \perp} < \chi^{\,\,\mu}_{\mu\,\, ||}$.
\end{itemize}
The same interpretation in terms of the mean free path for the Gauss-Bonnet term can be applied here. Considering the mean free path in the anisotropic direction $\ell_{\text{mfp}}^{||}$ and in the transverse plane $\ell_{\text{mfp}}^{\perp}$, we note that the mean free path of an anisotropic system in the transverse plane is larger than the corresponding quantity in the anisotropic direction, because $\eta_{\perp} > \eta_{||}$. This can be associated with a smaller drag force, a smaller jet quenching parameter, less screening (larger screening length), a smaller drag force and less photon production in the transverse plane when compared with the corresponding quantities in the anisotropic direction. 

It would be interesting to see how these observables behave for similar models. As far as we are aware, the only model that incorporates both the anisotropy and the higher curvature correction is \cite{Bhattacharyya:2014wfa}. Possible extensions of this work include the study of how the anisotropy and the higher derivative terms affect other observables like the imaginary part of the quarkonium potential, the quarkonium dissociation length in a plasma wind, Langevin diffusion coefficients, the dilepton production rate or the elliptic flow of photons and dileptons, to name a few.

\section*{Acknowledgements}
It is a pleasure to thank Diego Trancanelli for helpful discussions and insightful comments on the draft.
We are supported by FAPESP grants 2014/07840-7(ASM) and 2014/01805-5 (VJ).

\appendix
\section{Drag force for a general background and arbitrary direction} \label{app:drag}

In this appendix we derive a formula for the drag force. The metric background is assumed to be of the form
\begin{equation} \label{eq:arbitraryback}
 ds^2=G_{tt}dt^2+G_{xx}(dx^2+dy^2) + G_{zz}dz^2+G_{uu}du^2.
\end{equation}
We will only assume rotational symmetry in the $x y$ directions and consider the metric to depend only on $u$. This is essentially what was done in \cite{Giataganas:2012zy}, but here we consider the motion of the quark along an arbitrary direction, as in \cite{Chernicoff:2012iq}. 

The Nambu-Goto action for the string is given by
\begin{equation} \label{eq:NG}
 S=-\frac{1}{2\pi \alpha'}\int d\tau d\sigma \, e^{\phi/2}\sqrt{-\det g}=\int d\tau d\sigma \mathcal{L},
\end{equation}
where $\phi=\phi(u)$ is the dilaton field. By rotational symmetry in the $x y$ directions we can set $y=0$. We choose static gauge $(t,u)=(\tau,\sigma)$ and let us consider the motion of the quark in the $x z$ plane with a string embedding
\begin{equation}
 x(t,u)=(v t + \xi(u)) \sin\varphi,\qquad z(t,u)=(v t + \zeta(u))\cos\varphi,
\end{equation}
where $\varphi$ is the angle of the quark trajectory with the $z$ axis, i.e., $\varphi=0$ corresponds to the motion parallel with the anisotropic direction and $\varphi=\pi/2$ corresponds to the motion in the transversal direction. The boundary conditions are $\xi(u\to0)=0$ and $\zeta(u\to0)=0$, which are necessary in order to reproduce the stationary motion of the quark.

First, we need to compute the induced metric $g_{\alpha\beta}=G_{\mu\nu}\partial_\alpha x^\mu \partial_\beta x^\nu$ on the string worldsheet,
\begin{equation}
 g_{\alpha\beta}=\left(
\begin{array}{cc}
 G_{tt}+v^2 (G_{zz}  \cos ^2\varphi+G_{xx} \sin ^2\varphi) & v \left(G_{zz} \zeta
   '(u) \cos ^2\varphi+G_{xx} \xi ' \sin ^2\varphi \right) \\
 v \left(G_{zz} \zeta ' \cos ^2\varphi+G_{xx} \xi ' \sin ^2\varphi \right) &
   G_{uu}+G_{zz} \zeta '^2 \cos ^2\varphi+G_{xx}  \xi '^2\sin ^2\varphi \\
\end{array}
\right),
\end{equation}
where the prime denotes the derivative w.r.t. $u$. Ignoring factors of $\frac{1}{2\pi\alpha'}$, the Lagrangian takes the form
\begin{align}
 \mathcal{L} & = -e^{\phi/2}\left[-G_{zz} \cos^2\varphi (\zeta^{\prime2} G_{tt}+G_{uu} v^2+G_{xx} v^2 (\zeta'-\xi')^2\sin^2\varphi)-\right.\nonumber\\
 &\hspace{0.4 \linewidth}\left. -G_{xx} \sin^2\varphi(G_{tt} {\xi'}^2+G_{uu} v^2)-G_{tt} G_{uu}\right]^{\frac{1}{2}}.
\end{align}
We have the associated (conserved) momentum flow
\begin{align}
 \Pi_x & =\frac{\delta\mathcal{L}}{\delta x'} =-\frac{e^{\phi}G_{xx} \sin \varphi }{\mathcal{L}} \left(G_{tt} \xi'-G_{zz} v^2 (\zeta'-\xi') \cos^2\varphi\right),\\
 \Pi_z & =\frac{\delta\mathcal{L}}{\delta z'}= -\frac{e^{\phi}G_{zz} \cos \varphi }{\mathcal{L}} \left( G_{tt}\zeta'+G_{xx} v^2 (\zeta'-\xi') \sin ^2\varphi\right).
\end{align}
The values of the momenta will be fixed by imposing the requirement that $\xi'$ and $\zeta'$ are both real. To do this, we invert the above expression, writing
\begin{equation}
 \xi'(u)=\frac{G_{zz}N_x}{G_{xx}N_z}\zeta'(u),
\end{equation}
where we have defined the quantities
\begin{align}
 N_x & = G_{tt} \Pi_x \csc (\varphi )+G_{xx} v^2 (\Pi_x \sin (\varphi)+\Pi_z \cos (\varphi )),\\
 N_z & = G_{tt} \Pi_z \sec (\varphi )+G_{zz} v^2 (\Pi_x \sin (\varphi)+\Pi_z \cos (\varphi )).
\end{align}
Then we can use, for example, the expression for $\Pi_z$ to obtain $\zeta'$. The final expressions we found are given by
\begin{equation}
 \xi'=\pm\sqrt{-\frac{2G_{uu}G_{zz}}{G_{tt}G_{xx}D}}N_x,  \qquad 
 \zeta'=\pm \sqrt{-\frac{2G_{uu}G_{xx}}{G_{tt}G_{zz}D}}N_z,
\end{equation}
where 
\begin{align}
 D & = 2G_{tt} \left(G_{xx} \Pi_z^2+G_{zz} \Pi_x^2\right)
 +G_{xx} G_{zz}\left[G_{tt} e^{\phi}\left(2G_{tt}+v^2 (G_{xx}+G_{zz})\right)+ v^2 \left(\Pi_x^2+\Pi_z^2\right)\right]+\nonumber\\
   & \qquad +G_{xx}G_{zz} v^2 \left[ \left(-G_{tt} (G_{xx}-G_{zz}) e^{\phi}-\Pi_x^2+\Pi_z^2\right)\cos (2 \varphi )+2 \Pi_x \Pi_z \sin (2 \varphi )\right].
\end{align}
There is a sign ambiguity here, which we will fix later. The condition that \(\xi '\) and \(\zeta '\) are always real can be satisfied only if $D$ is positive for all $u$. But, in general, \(D\) has two zeros (turning points). Thus, in order to satisfy the positivity condition the two zeros should coincide at some critical point \(u_c\). Also, the numerators \(N_x\) and \(N_z\) should vanish at the same point \(u_c\), because otherwise $\xi' $ and $\zeta' $ would diverge. We begin the analysis finding the zeros of the numerator.
Imposing $N_x$ and $N_z$ to vanish at $u_c$ gives us a relation between \(\Pi _x\) and \(\Pi _z\), 
\begin{align}
 \frac{\Pi_x}{\Pi_z}=\frac{G_{xx}}{G_{zz}}\tan\varphi\Big|_{u=u_c}.
\end{align}
Using this relation, we can find the two zeros of $D$ at \(u_c\). This gives us two equations; the first one is
\begin{equation}
 \left[\frac{2G_{tt}}{v^2}+G_{xx}+G_{zz}+(G_{zz}-G_{xx})\cos(2\varphi)\right]_{u=u_c}=0,
\end{equation}
which can be used to fix the value of the critical point $u_c$. The second equation completely fixes the values of \(\Pi _x\) and \(\Pi _z\) and gives us the drag force
\begin{equation}
 \Pi_x= e^{\phi/2} G_{xx}v\sin\varphi\Big|_{u=u_c},\qquad  \Pi_z= e^{\phi/2} G_{zz}v\cos\varphi\Big|_{u=u_c}.
\end{equation}
We have fixed the sign of the momenta to be positive, thus $\xi'$ and $\zeta'$ are both negative, which is consistent with the physical condition that the string ``trails'' behind the quark \cite{Gubser:2006bz,Herzog:2006gh} and not in front of it. Two special cases are obtained from (\ref{eq:dragformula}) by setting $\varphi=0$ and $\varphi=\pi/2$. This gives us the drag force parallel and transversal to the direction of motion of the quark, respectively,
\begin{equation}
 F_\mt{drag}^{\,||}= e^{\phi/2} G_{zz}v\Big|_{u=u_c} ,\qquad  
 F_\mt{drag}^{\perp}= e^{\phi/2} G_{xx}v\Big|_{u=u_c}.
\end{equation}

\section{Jet quenching parameter for an arbitrary motion} \label{app:jet}

In this appendix we derive a formula for $\hat{q}$ considering a motion in an arbitrary direction and generic background. The steps are basically the same of \cite{Chernicoff:2012gu}, but here the computation is carried out in Einstein frame and the metric is left generic, with only a few assumptions, which we will specify below.

We assume a five-dimensional background displaying rotational symmetry in the $xy$ directions, 
\begin{equation}
 ds^2=G_{tt}dt^2+G_{xx}(dx^2+dy^2)+G_{zz}dz^2+G_{uu}du^2.
\end{equation}
From the rotational symmetry we can choose the direction of motion within the $xz$-plane. We define rotated coordinates 
\begin{align}
 z & = Z\cos\theta-X\sin\theta, \nonumber \\
 x & = Z\sin\theta+X\cos\theta, \nonumber \\
 y & = Y.
\end{align}
The new coordinates $(X,Y,Z)$ are obtained from the old coordinates $(x,y,z)$ by a rotation of an angle $\theta$ around the $y$-axis. We choose $Z$ to be the direction of motion of the quark. The direction of the momentum broadening takes place in the $XY$-plane and it forms an angle $\varphi$ with the $Y$-axis. The prescription instructs us to consider a string with an endpoint moving at the speed of light along the $Z$ direction. The other  endpoint is separated by a small distance $\ell$ along the direction of the momentum broadening. Thus we have a string worldsheet whose boundary is a rectangular light-like Wilson loop with sizes $L^-$ (along the $Z^-$ direction) and $\ell$, where $L^-$ is assumed to be very large. Our task is to find a string worldsheet that minimizes the Nambu-Goto action satisfying this boundary condition. We then need to evaluate the on-shell Nambu-Goto action and expand it to second order in $\ell$ to obtain
\begin{equation}
 \vev{W^A(\mathcal{C})} \simeq \exp\left[-\frac{L^-\ell^2}{4\sqrt{2}}\hat{q}\right],
\end{equation}
from which we extract the jet quenching parameter. It is convenient to work in light-cone coordinates
\begin{equation}
 t=\frac{Z^-+Z⁺}{\sqrt{2}}, \quad Z=\frac{Z^--Z^+}{\sqrt{2}}.
\end{equation}
The metric in these new coordinates takes the form
\begin{equation}
G_{\mu\nu}^{(LC)}=
\left(
\begin{array}{rrrrr}
 G_{++} & G_{+-} & G_{X-} & 0 & 0 \\
 G_{+-} & G_{++} & -G_{X-} & 0 & 0 \\
 G_{X-} & -G_{X-} & G_{XX} & 0 & 0 \\
 0 & 0 & 0 & G_{xx} & 0 \\
 0 & 0 & 0 & 0 & G_{uu} \\
\end{array}
\right),
\end{equation}
where
\begin{align}
 G_{++} & = \frac{1}{2} \left(G_{tt}+G_{xx} \sin ^2\theta+G_{zz} \cos ^2\theta\right) ,\nonumber \\
 G_{+-} & = \frac{1}{2} \left(G_{tt}-G_{xx} \sin ^2\theta-G_{zz} \cos ^2\theta\right) ,\nonumber \\
 G_{X-} & = \frac{\sin \theta \cos \theta }{\sqrt{2}}(G_{xx}-G_{zz}), \nonumber \\
 G_{XX} & = G_{xx} \cos ^2\theta+G_{zz} \sin ^2\theta .
\end{align}
We choose the static gauge $(\tau,\sigma)=(Z^-,u)$. Since we are assuming $L^-$ to be very large, there is a translational symmetry in the $Z^-$ direction, and we can fix the string embedding to only depend on $u$,
\begin{equation}
 Z^+=Z^+(u), \quad X\to X(u) \sin\varphi, \quad  Y\to Y(u)\cos\varphi.
\end{equation}
The induced metric on the string worldsheet, $g_{\alpha\beta}=G_{\mu\nu}\,\partial_\alpha x^\mu\partial_\beta x^\nu$, is given by
\begin{align}
 g_{\tau\tau} & = G_{++}, \quad\quad
 g_{\tau\sigma} = \sin \varphi \,G_{X-} {X'}+G_{+-} {(Z^+)'},\nonumber \\
 g_{\sigma\sigma} & = G_{uu}+\sin^2\varphi \, G_{XX} {X'}^2-2 \sin \varphi \,G_{X-}
   {(Z^+)'} {X'}+\cos ^2\varphi \,G_{xx} {Y'}^2+G_{++} {(Z^+)'}^2,
\end{align}
where the primes denote the derivative w.r.t. $u$. We can now write the Nambu-Goto action,\footnote{The extra factor of 2 comes from the two branches of the string worldsheet.}
\begin{equation}
 S=-2\frac{L^-}{2\pi\alpha'}\int_0^{\uh} du \, e^{\phi/2} \sqrt{-\det g}\equiv\frac{L^-}{\pi\alpha'}\int_0^{\uh} du\, \mathcal{L},
\end{equation}
where $\phi=\phi(u)$ is the dilaton scalar field and
\begin{align}
 \mathcal{L} = & -e^{\phi /2} \left[(G_{+-}+G_{++})\left(2 G_{X-} {X'} {(Z^+)'} \sin \varphi -G_{XX} {(Z^+)'}^2\right)\right. \nonumber \\
 & \qquad\quad \left. -G_{++} \left(G_{uu}+G_{xx} {Y'}^2 \cos ^2\varphi \right)+{X'}^2 \sin^2\varphi \left(G_{X-}^2-G_{++} G_{XX}\right)\right]^{\frac{1}{2}}.
\end{align}
Since the Lagrangian does not depend on $Z^+$, $X$, and $Y$, we have three conserved quantities, given by the canonical conjugate momenta. Up to a constant factor, they are given by
\begin{align}
\Pi_{+} & = \frac{e^{\phi } }{\mathcal{L}}(G_{+-}+G_{++}) (G_{X-} {X'} \sin \varphi-G_{XX} {(Z^+)'}), \nonumber \\
\Pi_X & =\frac{e^{\phi } }{\mathcal{L}}\sin \varphi \left[G_{X-} {(Z^+)'} (G_{+-}+G_{++})+{X'} \sin \varphi\left(G_{X-}^2-G_{++} G_{XX}\right)\right], \nonumber \\
 \Pi_Y & = -\frac{e^{\phi } }{\mathcal{L}}G_{++} G_{xx} {Y'} \cos ^2\varphi.
\end{align}
We are interested in the limit where $\Pi_+$, $\Pi_X$, and $\Pi_Y$ are small.\footnote{This is a consequence of $\ell$ be small, as explained in \cite{Chernicoff:2012gu}.}
Working in first order in the $\Pi_+$, $\Pi_X$ and $\Pi_Y$, we can invert the above expressions to obtain $(Z^+)', X'$ and $Y'$, we find that
\begin{align}
 {(Z^+)}' & = c_{++} \Pi_{+}+c_{+X} \Pi_X \csc \varphi, \nonumber \\
 X' & = c_{X+} \Pi_{+} \csc \varphi+c_{XX} \Pi_X \csc ^2\varphi, \nonumber \\
 Y' & = c_{YY} \Pi_Y \sec ^2\varphi,
\end{align}
where
\begin{align}
c_{++} & = \frac{e^{-\phi /2} G_{uu} \left(G_{X-}^2-G_{++} G_{XX}\right)}{(G_{+-}+G_{++}) \sqrt{-G_{++} G_{uu}} \left(G_{XX}^2-2 G_{X-}^2\right)} ,\nonumber \\
c_{+X} & = c_{X+}= \frac{e^{-\phi /2} G_{uu} G_{X-}}{\sqrt{-G_{++} G_{uu}} \left(2 G_{X-}^2-G_{XX}^2\right)} ,\nonumber \\
c_{XX} & = -\frac{e^{-\phi /2} G_{uu} G_{XX}}{\sqrt{-G_{++} G_{uu}} \left(G_{XX}^2-2 G_{X-}^2\right)} ,\nonumber \\
c_{YY} & = -\frac{e^{-\phi /2} G_{uu}}{G_{xx} \sqrt{-G_{++} G_{uu}}}.
\end{align}
Integration of ${Z^+}'$ gives zero. Integration of $X'$ gives $\ell/2$. Integration of $Y'$ also gives $\ell/2$. The conclusion is that
\begin{align}
\Pi_{+} & = \frac{\ell \sin \varphi \left(\int_0^{\uh} c_{+X}(u) \, du\right)}{2 \left(\left(\int_0^{\uh}c_{+X}(u) \, du\right){}^2-\left(\int_0^{\uh} c_{++}(u) \, du\right) \int_0^{\uh} c_{+X}(u) \, du\right)}\,, \nonumber \\
\Pi_X & = \frac{\ell \sin ^2\varphi \left(\int_0^{\uh} c_{++}(u) \, du\right)}{2 \left(\int_0^{\uh} c_{++}(u) \, du\right) \int_0^{\uh} c_{+X}(u) \, du-2 \left(\int_0^{\uh} c_{+X}(u) \, du\right){}^2}\,,\nonumber \\
\Pi_Y & = \frac{\ell \cos ^2\varphi}{2 \int_0^{\uh} c_{YY}(u) \, du}\,.
\end{align}
The on-shell action then takes the form, up to second order in the momenta,
\begin{equation}
 S=2i\frac{\sqrt{\lambda}L^-}{4\pi}\int_{0}^{\uh}du\left[c_{++}\Pi_+^2+\frac{1}{\sin^2\varphi}c_{XX}\Pi_X^2+\frac{2}{\sin\varphi}c_{+X}\Pi_+ \Pi_X+\frac{1}{\cos^2\varphi}c_{YY}\Pi_Y^2\right].
\end{equation}
Using the expressions for the coefficients the action can be rewritten as
\begin{equation}
 S=2i\frac{\sqrt{\lambda}L^-\ell^2}{16\pi}\left(\hat{P}(\theta)\sin^2\varphi+\hat{Q}(\theta)\cos^2\varphi\right),
\end{equation}
where
\begin{align}
 \hat{P}(\theta) & \equiv \frac{ \int_0^{\uh} c_{++}(u) \, du}{ \left(\int_0^{\uh} c_{++}(u) \, du\right) \int_0^{\uh} c_{+X}(u) \, du- \left(\int_0^{\uh} c_{+X}(u) \, du\right){}^2}\,, \nonumber \\
 \hat{Q}(\theta) & \equiv \frac{1}{\int_{0}^{\uh}c_{YY}du}\,.
\end{align}
From this we immediately extract the jet quenching parameter
\begin{equation}
 \hat{q}=\frac{\sqrt{2\lambda}}{\pi}\left(\hat{P}(\theta)\sin^2\varphi+\hat{Q}(\theta)\cos^2\varphi\right).
\end{equation}

\section{Quarkonium static potential in generic background}\label{app:Vqq}

In this appendix we derive a formula for the static potential of a quark-antiquark pair (quarkonium).\footnote{This computation is similar to what was done in \cite{Giataganas:2012zy}, generalizing the prescription of \cite{qqSonnen} for an anisotropic background.} Let $L$ be the separation between the quarks and assume a generic background of the form (\ref{eq:arbitraryback}). The dual picture corresponds to an U-shaped open string whose endpoints are located at the boundary and are identified with the quarks. Our task is to find the string worldsheet that minimizes the Nambu-Goto action (\ref{eq:NG}). By rotational symmetry in the $xy$-plane we can assume the pair to live in the $xz$-plane. Putting the center of mass of the pair at the origin, let $q$ be the polar radial coordinate and $\theta$ the angle between the pair and the $z$ direction. We fix the gauge $(\tau,\sigma)=(t,q)$. In this way the string embedding has the form
\begin{equation}
 X^\mu=(\tau,\sigma\sin\theta,0,\sigma\cos\theta,u(\sigma))
\end{equation}
The induced metric on the string worldsheet is given by
\begin{equation}
 g_{\tau\tau}=G_{tt}, \quad g_{\sigma\sigma}=G_{pp}+G_{uu}u'^2, \quad g_{\tau\sigma}=0,
\end{equation}
where $G_{pp}\equiv G_{zz}\cos^2\theta+G_{xx}\sin^2\theta$ and the prime denotes derivative w.r.t. $\sigma$. The Nambu-Goto action takes the form
\begin{equation}
 S=-\frac{\mathcal{T}}{2\pi\alpha'}\int_{-L/2}^{L/2}d\sigma \,e^{\phi /2} \sqrt{-G_{tt} \left(G_{pp}+G_{uu} u'^2\right)}\equiv \int_{-L/2}^{L/2}d\sigma\, \mathcal{L}.
\end{equation}
Since the Lagrangian $\mathcal{L}$ does not depend explicitly on $\sigma$, the Hamiltonian is a constant of motion
\begin{equation} \label{eq:H1}
H=\frac{\partial\mathcal{L}}{\partial \sigma'}\sigma'-\mathcal{L} =  -\frac{\mathcal{T}}{2\pi\alpha'}\frac{e^{\phi /2} G_{tt} G_{pp}}{\sqrt{-G_{tt} \left(G_{pp}+G_{uu} u'^2\right)}}.
\end{equation}
Evaluating the Hamiltonian at the turning point $u_0\equiv u(0)$, where $u'=0$, we find the value of the constant to be
\begin{equation} \label{eq:H2}
 C=\frac{\mathcal{T}}{2\pi\alpha'}e^{\frac{\phi}{2}} \sqrt{-G_{pp} G_{tt}}\Big|_{u=u_0}\,.
\end{equation}
In order to simplify the expressions, it is useful  to define the auxiliary quantities
\begin{equation}
 P\equiv e^{\frac{\phi}{2}} \sqrt{-G_{pp} G_{tt}}\,, \qquad
 Q\equiv e^{\frac{\phi}{2}} \sqrt{-G_{tt} G_{uu}}\,.
\end{equation}
Using (\ref{eq:H1}) and(\ref{eq:H2}) we can find an expression for $u'$,\footnote{One needs to be careful with the choice of sign here: when $\sigma$ goes from $0$ to $L/2$, then $u$ goes from $u_0$ to $0$ and thus $u'<0$.}
\begin{equation}
u'= \pm\frac{P}{Q}\frac{\sqrt{P^2-P_0^2}}{P_0}, \quad P_0\equiv P|_{u=u_0}.
\end{equation}
Integrating the above expression we find that the separation between the quarks is given by
\begin{equation}
 L=-2\int_{0}^{u_0}\frac{d\sigma}{du}du=2\int_{0}^{u_0}\frac{Q}{P}\frac{ P_0 }{ \sqrt{P^2-P_0^2}}\,.
\end{equation}
Before we compute the on-shell Nambu-Goto action to find the potential energy that keeps the pair bounded, we need to take care of the ultraviolet divergence due to the infinite mass of the quarks. The mass term corresponds to a string hanging down straight from the boundary to the horizon. Note that in this case the string goes from $0$ to $\uh$ while $\sigma$ is fixed, thus it effectively corresponds to $u'\to\infty$. Expanding the on-shell Nambu-Goto action in powers of $1/u'$ for this configuration we obtain 
\begin{equation}
 M_Q=-\frac{\mathcal{T}}{2\pi\alpha'}\int_{0}^{\uh}du\,Q +O\left(\frac{1}{u'}\right).
\end{equation}
Finally, computing the on-shell Nambu-Goto action for the U-shaped configuration with the mass subtraction we obtain the static potential
\begin{equation}
 V_{Q\bar{Q}}=\frac{S^{(\mt{on-shell})}}{\mathcal{T}}-2M_Q=-\frac{1}{2\pi\alpha'}\left[P_0 L + 2\int_{0}^{u_0}du\frac{Q}{P}\left(\sqrt{P^2-P_0^2}-P\right)-2\int_{u_0}^{\uh}du\,Q\right].
\end{equation}


\end{document}